\documentclass[]{aa}
\usepackage[varg]{txfonts}
\usepackage{graphicx}
\usepackage{aecompl,color, amsmath}
\usepackage{siunitx}
\usepackage[normalem]{ulem} 

\DeclareSIUnit\erg{erg}

\newcommand{\ha}{H$\alpha$}
\newcommand{\hb}{H$\beta$}

\newcommand{\hii}{\ion{H}{ii}~}
\newcommand{\paperI}{Paper~\textsc{i}}
\def\msun{\hbox{M$_\odot$}}

\usepackage{supertabular,booktabs}

\graphicspath{ {figures/} }

\begin{document}

\title{Stellar feedback in M83 as observed with MUSE}
\subtitle{II. Analysis of the \ion{H}{ii} region population: ionisation budget and pre-SN feedback}
\author{
Lorenza Della Bruna\inst{\ref{SU}} \and
Angela Adamo\inst{\ref{SU}} \and
Anna~F. McLeod\inst{\ref{durham1}, \ref{durham2}} \and
Linda~J. Smith\inst{\ref{space-telescope-esa}} \and
Gabriel Savard\inst{\ref{UL}} \and 
Carmelle Robert\inst{\ref{UL}} \and
Jiayi Sun\thanks{CITA National Fellow} \inst{\ref{mcmaster},\ref{toronto}, \ref{ohio}} \and 
Philippe Amram\inst{\ref{marseille}} \and
Arjan Bik\inst{\ref{SU}} \and
William~P. Blair\inst{\ref{johns-hopkins}} \and
Knox~S. Long\inst{\ref{space-telescope}} \and
Florent Renaud\inst{\ref{lund}}\and
Rene Walterbos\inst{\ref{new-mexico}}\and
Christopher Usher\inst{\ref{SU}} 
}

\institute{
The Oskar Klein Centre, Department of Astronomy, Stockholm University, AlbaNova, SE-10691 Stockholm, Sweden\label{SU} \and
Centre for Extragalactic Astronomy, Department of Physics, Durham University, South Road,  Durham DH1 3LE, UK\label{durham1}
\and
Institute for Computational Cosmology, Department of Physics, University of Durham, South Road, Durham DH1 3LE, UK\label{durham2}
\and
Space Telescope Science Institute, 3700 San Martin Drive, Baltimore MD 21218, USA\label{space-telescope-esa}
Département de physique, de génie physique et d'optique, Université Laval\label{UL}, 
\and
Department of Physics and Astronomy, McMaster University, 1280 Main Street West, Hamilton, ON L8S 4M1, Canada\label{mcmaster}
\and
Canadian Institute for Theoretical Astrophysics (CITA), University of Toronto, 60 St George Street, Toronto, ON M5S 3H8, Canada\label{toronto}
\and
Department of Astronomy, The Ohio State University, 140 West 18th Avenue, Columbus, OH 43210, USA\label{ohio}
\and
Aix-Marseille Université, CNRS, CNES, LAM, Marseille, France\label{marseille}
\and
The William H. Miller III Department of Physics and Astronomy, Johns Hopkins University, 3400 N. Charles Street, Baltimore, MD 21218, USA\label{johns-hopkins}
\and 
Space Telescope Science Institute, 3700 San Martin Drive, Baltimore MD 21218, USA; Eureka Scientific, Inc. 2452 Delmer Street, Suite 100, Oakland, CA 94602-3017, USA\label{space-telescope}
\and
Department of Astronomy and Theoretical Physics, Lund Observatory, Box 43, 221 00 Lund, Sweden\label{lund}
\and
Department of Astronomy, New Mexico State University, Las Cruces, NM, 88001, USA\label{new-mexico}
}

\authorrunning{Della Bruna et al.}
\titlerunning{Stellar feedback in M83}

\date{Received 22 February 2022 / Accepted 17 June 2022}

\abstract
{
Energy and momentum injected by young, massive stars into the surrounding gas play an important role in regulating further star formation and in determining the galaxy's global properties. Before supernovae begin to explode, stellar feedback consists of two main processes: radiation pressure and photoionisation.
}
{
We study pre-supernova feedback and constrain the leakage of Lyman continuum (LyC) radiation in a sample of $\sim$ 4700 \hii regions in the nearby spiral galaxy M83. We explore  the impact that the galactic environment and intrinsic physical properties (metallicity, extinction, stellar content) have on the early phases of \hii region evolution. 
}
{
We combine VLT/MUSE observations of the ionised gas with young star cluster physical properties derived from HST multiwavelength data. We identify \hii regions based on their \ha\ emission, and cross-match the sample with planetary nebulae and supernova remnants to assess contaminant sources and identify evolved \hii regions. We also spectroscopically identify Wolf-Rayet (WR) stars populating the star-forming regions.
We estimate the physical properties of the \hii regions (luminosity, size, oxygen abundance and electron density). For each \hii region, we compute the pressure of ionised gas ($P_{\rm ion}$) and the direct radiation pressure ($P_{\rm dir}$) acting in the region, and investigate how they vary with galactocentric distance, with the physical properties of the region, and with the pressure of the galactic environment ($P_\mathrm{DE}$). For a subset of $\sim$ 500 regions, we also investigate the link between the pressure terms and the properties of the cluster population (age, mass and LyC flux).
By comparing the LyC flux derived from \ha\ emission with the one modelled from their clusters and WRs, we furthermore constrain the escape of LyC radiation from the \hii regions ($f_{\rm esc}$).
}
{
We find that $P_{\rm ion}$ dominates over $P_{\rm dir}$ by at least a factor of 10 on average  over the disk. Both pressure terms are strongly enhanced and become almost comparable in the central starburst region. In the disk ($R \geq 0.15~R_e$), we observe that $P_{\rm dir}$ stays approximately constant with galactocentric distance. $P_{\rm dir}$ is positively correlated with an increase in radiation field strength (linked to the negative metallicity gradient in the galaxy), while it decreases in low extinction regions, as expected if the amount of dust to which the momentum can be imparted decreases. $P_{\rm ion}$ decreases constantly for increasing galactocentric distances; this trend correlates with the decrease in extinction -- indicative of more evolved and thus less compact regions -- and with changes in the galactic environment (traced by a decrease in $P_\mathrm{DE}$). In general, we observe that \hii regions near the center are underpressured with respect to their surroundings, whereas regions in the rest of the disk are overpressured and hence expanding.
We find that regions hosting younger clusters or having more mass in young star clusters have a higher internal pressure, indicating that clustered star formation is likely playing a dominant role in setting the pressure.
Finally, we estimate that only 13\% of \hii regions hosting young clusters and WR stars have $f_{\rm esc} \geq 0$, which suggests that star formation taking place outside young clusters makes a non-negligible contribution to ionising \hii regions.
}
{}
\maketitle 

\section{Introduction}
\label{section:introduction}
Stellar feedback consists of a variety of processes~\citep[see][for a review]{krumholz14_review,dale15}, the most important mechanisms being photoionisation, direct radiation pressure, and mechanical feedback via stellar winds and supernovae (SNe) explosions. The combined effect of these mechanisms results in a multi-scale phenomenon, ranging from scales of a few parsec -- surrounding the stars -- to galactic-wide scales.

 Stellar feedback originates from massive stars, forming in the densest cores of giant molecular clouds (GMCs). Therefore, to study stellar feedback and its regulatory role in the star formation cycle of galaxies, it requires access to a large dynamical range of observations and simulations that capture processes happening over five orders of magnitude in physical scales. One of the key questions currently focuses on the timescales necessary to dissolve GMCs. These timescales are the fundamentals because they determine the resulting efficiency of the star formation process in the region, as well as how energy and momentum stream away from these regions maintaining a multi-phase ISM.

Numerical approaches typically focus on simulating isolated star-forming regions, but including detailed treatment of star formation and stellar feedback \citep[e.g.][among the latest]{Kim2018, kim2021, Olivier2021, grudic2021, grudic2022}; or probe the feedback in isolated galaxy simulations by focusing on different feedback processes, while simplifying other physical processes happening at small physical scales \citep[e.g.][]{hopkins2018, bending20, jeffreson2021}; or re-simulate regions of galaxies (e.g. a fraction of spiral arms) to preserve the regulatory role of galactic scale dynamics, while improving the details of feedback prescriptions \citep[e.g.][]{gatto2017, ali21, ali2022, bending2022}. Overall, these diverse approaches reach similar conclusion regarding the importance that photionisation from massive stars has in the evolution of the star-forming regions, by lowering the gas density and, therefore, pre-processing the surrounding gas where supernovea (SNe) will explode.

From the observational side, great advancements have recently been done thanks to the advent of sensitive  integral field spectrographs (IFS) with wide field of views, enabling to cover large portions of local galaxies at reasonable high spatial resolution. It is now possible to directly study the impact of different types of feedback on the star-forming regions, and to trace their rapid evolution. Two instruments that have been playing an important role in this sense are: the MUSE IFS \citep{bacon10} at ESO's Very Large Telescope and the SITELLE imaging fourier transform spectrograph at the Canada-France-Hawaii Telescope \citep{drissen19}.
Using these instruments, two large ongoing surveys targeting \hii regions in nearby galaxies at scales relevant for these types of studies are the PHANGS-MUSE \citep{emsellem22} and SITELLE-SIGNALS \citep{rousseau19} surveys, which are mapping, respectively, $\sim$ 20 and 40 nearby galaxies at a median physical scale of 50 pc. These surveys are providing us with a statistical sample of \hii regions, enabling us to study their overall properties such as luminosity, metallicity, ionisation state and indirectly derive their internal pressure, and how they depend on the galactic environment (e.g. galactocentric distance or arm/interarm environment), on changes in local environmental conditions and on the average properties of the stellar populations hosted by the regions \citep[e.g.][]{rousseau18, kreckel19, kreckel20, barnes21}.  These studies,  however, do not allow to resolve sizes and determine electron densities for a large fraction of their \hii regions, requiring indirect methods and assumptions to estimate different pressure terms and resulting in degeneracies not easy to disentangle \citep[e.g.][]{barnes21, barnes2022}. 
\par On the other hand, very high-resolution ($\sim 10$ pc scale) studies of smaller samples of \hii regions are allowing us to resolve the star-forming regions in their details. This makes possible investigations of the impact of the different stellar feedback mechanisms on individual regions \citep[e.g.][]{lopez14, mcleod19, mcleod20, mcleod21}, and how they are related to the properties of the regions (e.g. metallicity or extinction) and of their environment. If the stellar population of the regions is accessible, by modeling the expected ionising photon flux $Q(\mbox{H}^0)$ and comparing it to the observed ionised gas emission, one can infer whether the regions are leaking hydrogen ionising radiation \citep[Lyman continuum photons, LyC, h$\nu > 13.6$ eV; e.g.][]{mcleod19, mcleod20, dellabruna21}. By constructing a `ionisation budget' for the full sample, one can assess whether ionising photons escaping from the \hii region population can explain the amount of diffuse ionised gas (DIG) emission outside the \hii regions, shedding light on the origin of this emission component of the interstellar medium \citep[ISM, see e.g. the review of][]{haffner09}.
\par Another open question is how the LyC escape fraction ($f_{\rm esc}$) is linked to the properties of the regions, such as their ionisation structure (e.g. the presence of optically thin `channels'), or the stellar population they host. Recent high-resolution cosmological simulations of \citet{ma20} seem to indicate for example that regions with an age spread in their stellar population are advantaged in leaking ionising photons. Namely, feedback from clustered SNe can result in the creation of a superbubble; a second generation of stars is then able to ionise pre-cleared lower density channels, and leak LyC photons into the surrounding ISM. In a previous publication \citep{dellabruna21} we investigated this in a sample of 8 \hii regions in the nearby galaxy NGC~7793, finding a significant leakage of ionising photons but no conclusive evidence of a trend with age spread.
\par In this work, we study a sample of $\sim$ 4700 \hii regions across the stellar disk ($R \lesssim 1.1~R_e$) of the nearby galaxy M83, a grand design barred spiral at a distance $\simeq$ 5 Mpc \citep[][see Table~\ref{table:m83_param}]{jacobs09}.
In \citet[][henceforth \paperI]{dellabruna22} we have presented a large MUSE mosaic of M83 (3.8 $\times$ 3.8 kpc), with a spatial resolution of 20 pc.
In \paperI, we discussed the large scale kinematics of the gas and the stars. Here, we focus on the individual \hii regions. With the spatial resolution of our data, we are able to resolve most individual regions. We have access to their stellar population thanks to HST observations of the young star cluster (YSC) population \citep{silva-villa14, adamo15}. We will investigate the relative importance of different feedback mechanisms and constrain the escape of ionising radiation from the regions. We then investigate how these quantities are linked to the region properties and the stellar population they host.

\par This work is organised as follows: in Sect.~\ref{section:data} we briefly describe the dataset. In Sect.~\ref{section:hii_regions} we summarise the selection steps and properties of the \hii regions sample. In Sect.~\ref{section:stellar_population} we describe the stellar population in the regions, and in Sect.~\ref{section:hii_properties} we summarise their physical properties. In Sect.~\ref{section:pressure} and \ref{section:budget} we investigate the contribution of different pressure terms in the regions and compute their ionisation budget. We discuss the results in Sect.~\ref{section:discussion}, and conclude with a summary in Sect.~\ref{section:conclusions}.

\section{Data description}
\label{section:data}

\begin{table}
\caption{Adopted parameters of M83.}
\begin{tabular}{lcc}
\hline \hline
Parameter & Value & Ref. \\ \hline
Distance & 4.89 Mpc ($1\arcsec = 24$ pc) & (1) \\
Effective radius ($R_e$) & 3.5 kpc & (2)\\
\hline
\end{tabular}
\tablebib{(1) \citet{jacobs09}; (2) \citet{leroy21}.}
\label{table:m83_param}
\end{table}

We summarise the main physical properties adopted for M83 in Table~\ref{table:m83_param}. The dataset is described in detail in \paperI. Briefly, we
 constructed a large mosaic of 26 MUSE pointings, combining a total of 65 single exposures\footnote{The data are part of the observing programs  096.B-0057(A) and 0101.B-0727(A) (PI Adamo, 46 exposures), 097.B-0899(B) (PI Ibar, 15 exposures) and 097.B-0640(A) (PI Gadotti, 4 exposures).}. The data are obtained in Wide-Field Mode (WFM), and extended wavelength mode (4650 -- 9300 \AA{}), and cover a galactocentric radius of $\sim 3.8$ kpc ($1.1 \times R_e$), for a total area of 40.5 kpc$^2$. The median point spread function (PSF) measured at 7000~\AA{} is 0\farcs7 (17 pc); we refer to \paperI\ for details on PSF variation with wavelength and across the mosaic tiles.

\par We combine the information on the ionised gas from MUSE with HST data tracing the star cluster population \citep{silva-villa14, adamo15}. M83 was first observed during the WFC3 Early release science program (GO11360, PI O'Connell). The coverage was later extended to a galactocentric radius of 4.5 kpc ($1.3 \times R_e$, GO12513, PI Blair). The final HST mosaic\footnote{Publicly available at \url{https://archive.stsci.edu/prepds/m83mos/}} (7 pointings) is described in \citet{blair14}, and the data have a FWHM of $0\farcs08$ (1.9 pc).


\begin{figure*}
\center
\includegraphics[width=19cm]{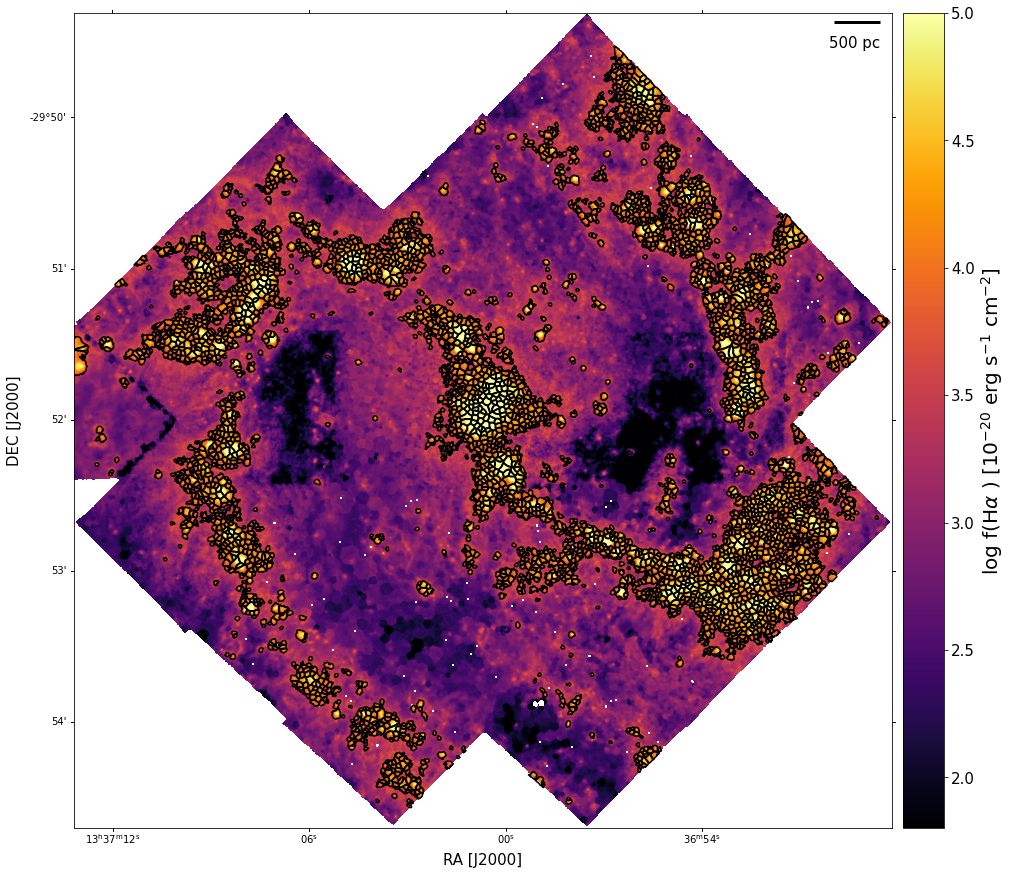}\\
\includegraphics[width=0.49\linewidth]{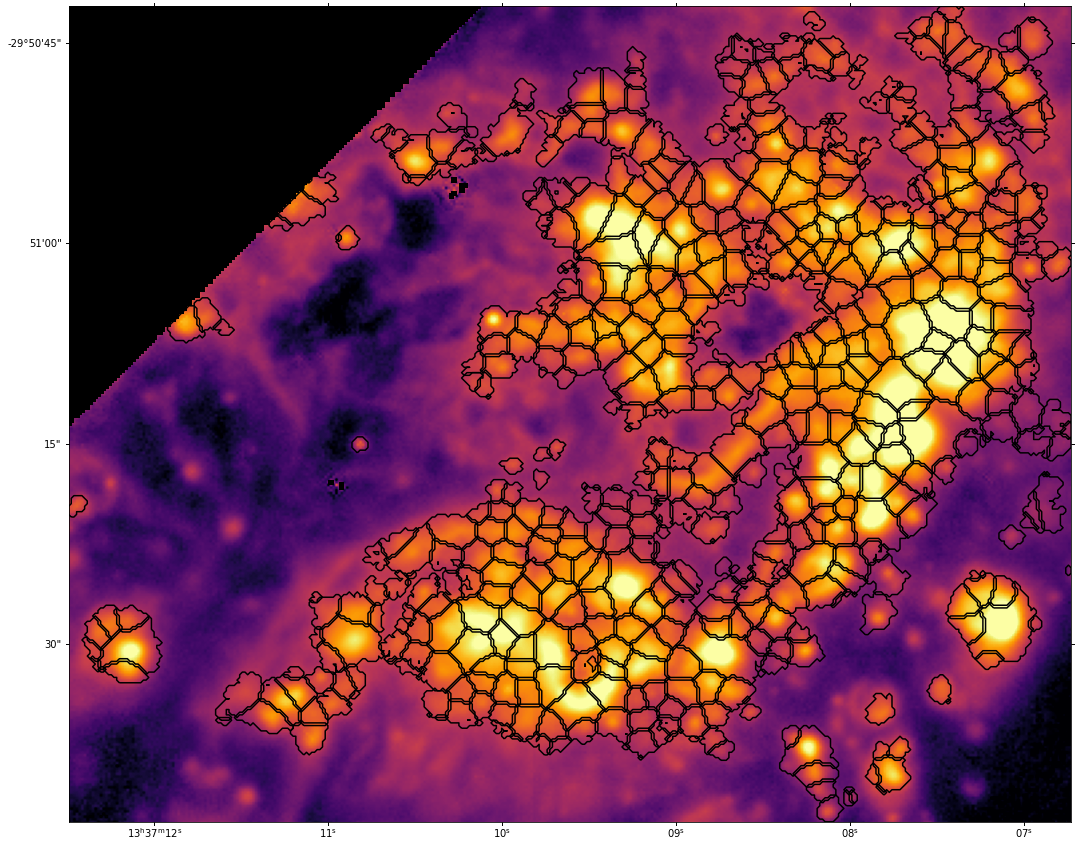}
\caption{Map of H$\alpha$ emission (extinction corrected) with the boundaries of the identified \hii regions. The bottom panel shows an enlargement of one of the star-forming complexes.}
\label{fig:hii_regions}
\end{figure*}

\section{\ion{H}{ii} regions, SNR, PNe sample}
\label{section:hii_regions}

\subsection{Identification of the \hii regions}
\label{section:hii_identification}
The identification of the boundaries of the \hii regions and other emission regions has been performed using the Python package \textsc{astrodendro}\footnote{\url{https://dendrograms.readthedocs.io}} as described in \paperI.
In summary, the dendrogram tree was computed down to a surface brightness (SB) of 1.23 $\times 10^{-15}$ erg s$^{-1}$ cm$^{-2}$ arcsec$^{-2}$. This threshold was estimated as the brightness of an \hii region of 10 pc radius ionised by a single low-luminosity star, adopting the luminosity
of a O9.5 main sequence star \citep{martins05}. We set the minimum leaf size of the tree to a square of width 4.3 pixel, corresponding to the typical FWHM of the PSF at \ha\ (see \paperI). The outcome of \textsc{astrodendro} is a tree structure, consisting of leaves and branches organised hierarchically according to their flux. In crowded \hii region complexes however, this is often not sufficient to disentangle single regions. This is less of an issue in flocculent spiral galaxies (NGC~7793, \citealp{dellabruna20}, or NGC~300, \citealp{mcleod21}), where \hii complexes are not too large, but can be a cause of concern in grand-design spirals such as M83, where the spiral arms consist of tightly packed \hii regions.

\par Breaking up \hii region complexes into single regions has therefore required additional steps.
We first tested the approach of \citet{mcleod21}, where hierarchical structures identified with \textsc{astrodendro} are divided into sub-structures using the \textsc{scimes} algorithm \citep{colombo15}. Hereby, relevant sub-structures are identified and grouped by determining their `affinity' with a spectral clustering approach. However, the \textsc{scimes} algorithm did not perform optimally for our dataset, due to the wide range in luminosity of the \hii regions and the lack of flexibility in the code parameters. Namely, the code starts by identifying sub-structures in the most luminous complexes. Increasing the \texttt{user\_k} parameter (expected number of clusters) only results in an over-shredding of the bright complexes, and a lack of substructures in the less luminous ones.

\par We obtain optimal results using the first version of the code developed by Savard et al. (in prep.) which is based on the algorithm described in \citet{rousseau18}. For details on the working principles of the code, we refer the reader to \citet{rousseau18} and the upcoming work of Savard et al, while we summarise here principal steps. The algorithm works as follows: in a \textit{first step}, peaks of emission are identified in a \ha\ linemap by computing a Laplacian map.
Relevant peaks of emission are identified using the \texttt{peak\_finder} function, which takes as input:
\begin{itemize}
    \item the size of a (square) detection box;
    \item the standard deviation of the Gaussian filter used by the Laplacian function to convolve the image.
    \item the coordinates of a `background' box, located in an area with little to no emission;
    \item a constant $f_{\rm noise}$, which determines the relative importance of the local noise variations and uncertainties on the \ha\ emission (see Eq.~\ref{eq:hii_threshold});
\end{itemize}
After detecting all relevant peaks of emission in the Laplacian image, the desired peaks are selected based on a detection threshold $t$. The latter is determined as a function of the background and peak emission level, as well as the local noise variation and the uncertainty on the emission peak in the detection box:
\begin{equation}
t = \big( m_{\rm detec} + m_{\rm bkg} + f_{\rm noise} \times \sqrt{\sigma_{\rm detec}^2 + \sigma_{\rm bkg}^2} \big) \times A_{\rm detec},
\label{eq:hii_threshold}
\end{equation}
where $m_{\rm detec}, m_{\rm bkg}$ and $\sigma_{\rm detec}, \sigma_{\rm bkg}$ are the median and standard deviation of the input map within the detection box and in the background box, respectively, and $A_{\rm detec}$ is the area of the detection box.
In a \textit{second step}, each spaxel is assigned to the peak minimizing $1/r^3$, where $r$ indicates the distance to the peak. This metric was found to prevent bright emission peaks from embedding spaxels physically linked to dimmer peaks.
\par As input to the code, we provide the (non continuum subtracted) map of the \ha\ emission, obtained by integration of the datacube around the \ha\ line in the (restframe) wavelength range 6559 -- 6569~\AA{}, and a `continuum map' obtained as the median of the data in the range around the line (stacking the regions 6529 -- 6539~\AA{} and 6599 -- 6609~\AA{}).
We used a peak detection box size of 2 pixels (limited by the seeing) and a background detection box size of $\sim 130 \times 160$ arcsec$^2$ situated in an interarm region with little \ha\ emission.
We set the standard deviation of the Laplacian filter to 1.5. This value has been determined by visually inspecting the location of the resulting peaks. A lower value results in the detection of noise peaks, whereas a larger one retains only the strongest peaks of emission.
We tested several value of $f_{\rm noise}$. By visual inspection, we found the optimal value to be around 1. Higher $f_{\rm noise}$ values lead to missed detections of relevant peaks of emission, resulting in irregularly shaped and unrealistic region boundaries. Lower $f_{\rm noise}$ values, on the other hand, do not impact the distribution of the peaks within the dendrogram contours, but only result in a deeper detection limit below the adopted minimum SB threshold. In Appendix~\ref{section:appendix_HII}, we illustrate an example of boundaries, luminosity and sizes of regions identified using different values of $f_{\rm noise}$. The change in $f_{\rm noise}$ affects the detection of faint \hii regions, while it does not affect the recovered size distributions. 
To define the final domain of the regions, we set a termination criterion by multiplication with the mask obtained from \textsc{astrodendro}.

\par We note that throughout this work we do not correct the \hii region emission for background emission caused e.g. by the DIG or other nearby regions. This emission varies significantly with distance from the regions, and is therefore very hard to estimate in crowded areas. DIG contamination by itself should not strongly affect the \ha\ luminosity, but can explain e.g. why some regions are beyond the extreme starburst line in the BPT diagrams in Fig.~\ref{fig:hii_bpt} and have a [\ion{S}{ii}]~$\lambda\lambda$6716,31 ratio above the sensitivity limit to the electron density \citep[e.g.][]{belfiore22}.
We also would like to point out that, despite our best efforts, not all regions will have well physically motivated boundaries in crowded areas. Nevertheless, the flux of each region will be dominated by the peak of \ha\ emission contained within it.

\par The resulting regions and an enlargement of a \hii region complex are shown in Fig.~\ref{fig:hii_regions}. We identify a total of 4687 candidate regions. However, other classes of objects such as supernova remnants (SNRs) and planetary nebulae (PNe) emit in \ha. In the following we therefore cross-match the \hii candidates with SNRs and PNe catalogues to remove such contaminants.

\subsection{Identification of SNR}
\label{section:snr}
We cross-match our \hii region catalogue with SNR identified in the MUSE M83 dataset by \citet{long2022} and based on the [\ion{S}{ii}]/\ha\ line ratio. The catalogue consists of 228 SNR in the region covered by the MUSE data. We find that 149 of our emission regions host a SNR (within a distance of 0.2\arcsec). We keep these in our \hii region sample but flag them as `evolved' \hii regions throughout our analysis.

\subsection{Identification of PNe}
\label{section:pn}

\begin{figure}
    \centering
    \includegraphics[width = \linewidth]{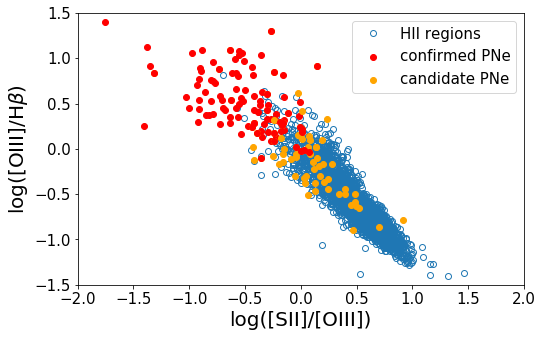}
    \caption{Location of spectroscopically confirmed PNe (filled red points), PNe candidates (filled orange points) and \hii regions (open blue circles) in a diagram of [\ion{O}{iii}]~$\lambda\lambda$4959,5007/\hb\ vs [\ion{S}{ii}]~$\lambda\lambda$6716,6731/[\ion{O}{iii}]~$\lambda\lambda$4959,5007. We applied a first order background correction to the PN fluxes as described in Sect.~\ref{section:pn}.
    We observe that the confirmed PNe occupy a well-defined region in the diagram.}
    \label{fig:o3hb_vs_s2o3_pn}
\end{figure}

PNe are emission nebulae arising from an old stellar population that we wish to exclude from our \hii region study. In the Milky Way and local galaxies, they are observed to have enhanced [\ion{O}{iii}]/\ha\ ratios \citep[see e.g.][]{ciardullo02,magrini05,kniazev08}. Here we use two different methods to identify them.

The first method makes use of the spectral information contained in the MUSE datacubes \citep[e.g.][]{kreckel2017, mcleod21}.  Using the (reddening corrected) [\ion{O}{iii}]/H$\alpha$ emission line map, 
we extract compact sources using \textsc{astrodendro}, requiring a minimum ratio value for [\ion{O}{iii}]/H$\alpha$ of 0.45 and a minimum number of 8 pixels per leaf.
Before extracting the regions of interest, we smooth the map with a 2D Gaussian filter with a kernel standard deviation of 1$\sigma$. This is done in order to prevent \textsc{astrodendro} from detecting noise in the map. We only consider leaves in the tree and obtain an initial sample of 3978 potential candidates.

Once the positions of the regions of interest are known, we extract spectra with a circular aperture of $r = $ 1\arcsec\, (5 px), using the leaf centres estimated from the \textsc{astrodendro} \texttt{PPStatistic} module, on the continuum-subtracted datacube. The spectra are dereddened using the E($B-V$) estimated from the Balmer decrement (see \paperI) at the center of the aperture. Confirmed candidates must then fulfill four criteria:
\begin{enumerate}
    \item Have f([\ion{O}{iii}]~$\lambda$5007)$_{aperture}$ $\geq$ 2 f([\ion{O}{iii}]~$\lambda$5007)$_{bkg}$, where the background flux is estimated in an annulus of $(r+4, r+6)$ pixels;
    \item Have a ratio of f([\ion{O}{iii}]~$\lambda$5007)/f(\ha) $>$ 0.5. We use this as a lower limit threshold; previously known PNe are generally observed to have much higher ratios \citep[$\sim$ 2,][]{ciardullo02};
   \item Have a ratio of f([\ion{S}{ii}]~$\lambda$6731)/f(\ha) $<$ 0.5. This is done in order to exclude SNR from the sample, which notoriously have a high ratio of [\ion{S}{ii}]/\ha\ \citep[$> 0.4$,][]{mathewson73}. We remark that here we use a less strict limit as a first filtering step, but that the candidates are later cross-checked with SNR identified based on a threshold of 0.4.
    \item Be unresolved in HST [\ion{O}{iii}] continuum subtracted imaging \citep{blair14}. The HST WFC3 pixel size of 0.04\arcsec\ corresponds to $\sim$1.1 pc, so PNe will remain point-like while most compact emission nebulae will be resolved. 
\end{enumerate}
We note that in criteria 2 and 3, the [\ion{O}{iii}]/\ha\ and [\ion{S}{ii}]/\ha\ ratios are background corrected by subtracting from each linemap a first order estimate of the emission background, computed over the entire FoV (via sigma clipping) rather than locally. This is done in order to avoid contamination by surrounding bright regions.

\par Of the initial 3978 candidates, 131 pass all four selection criteria. We cross-match this catalogue with the sample of SNR by \citet{long2022} and exclude 11 more candidates coincident with the location of a SNR (within 0.4\arcsec).
We are left with 120 confirmed PNe. Five of these were previously catalogued by \citet{herrmann09}, all located in the outer disk. Two additional candidates from \citet[][M83-100 and M83-195]{herrmann09} are located at the very edge of the MUSE FoV and are not picked up by our selection criteria.
The 115 new sources reported here highlight the capability of MUSE (combined with HST) in detecting this type of objects even at the distance of M83.
We inspected the confirmed PN sample for the presence of \ion{He}{ii}~$\lambda$4686; this line traces extremely high energetic photons with h$\nu \geq 54.4$ eV and, when detected, is a robust confirmation of the presence of a PN~\citep{frew10}. We find \ion{He}{ii} emission in 15 of the objects.
The position and coordinates of the 120 confirmed PNe can be found in Appendix~\ref{section:appendix_pn} (Fig.~\ref{fig:PN} and Table~\ref{table:PN}). 

The second method relies on the superior spatial resolution of the HST data. We use continuum subtracted [\ion{O}{iii}] and H$\alpha$ data \citep[][]{blair14} to create a line ratio map. The latter is visually inspected to identify potential PNe candidates that have been missed in the MUSE extraction above.  The selected candidates appear as point-like [\ion{O}{iii}] emission sources without stellar counterparts in the HST dataset. However, WFC3 data alone has some limitations. Faint stellar residuals in the subtracted emission line images and other artifacts such as cosmic ray residuals or hot camera pixels can be mistaken for real objects. MUSE data and HST WFC3 imaging used in conjunction are much more powerful for identifying PNe at the distance of M83 than either dataset is alone.

We visually inspected the candidates using the display program SAOimage ds9, simultaneously displaying the MUSE H$\alpha$, and [\ion{O}{iii}]/H$\alpha$ map alongside the subtracted WFC3 [\ion{O}{iii}] and the WFC3 V-band images during the search for potential candidates. Point-like sources in the WFC3 [\ion{O}{iii}] image are compared against the continuum band to remove stellar residuals from further consideration and against the aligned MUSE data to verify the presence of a corresponding emission nebula at the position, thus eliminating cosmic ray residuals.
We show an example of the search technique applied to a small 6\arcsec\ region of M83 in Fig.~\ref{fig:pn_visual_inspection} of Appendix \ref{section:appendix_pn}. Using this approach, we identify 124 new PNe candidates. We then extract the spectra of the candidates from the continuum subtracted MUSE datacube and estimate line ratios as described in the first method. Of the 124 candidates, 81 are detected in all the emission lines of interest (H$\alpha$, H$\beta$, [\ion{O}{iii}], [\ion{S}{ii}]). Of these 81 candidates, only 28 have f([\ion{O}{iii}]~$\lambda$5007)/f(\ha) ratios greater than 0.5. We do not require the f([\ion{S}{ii}]~$\lambda$6731)/f(\ha) $<0.5$ criterion as in these cases we use the prior knowledge of the SNR positions. These 28 candidates have therefore been included into the sample of spectroscopically confirmed PNe. The remaining sources are listed and referred to as candidates in Table~\ref{table:PN} and in Fig.~\ref{fig:PN} and Fig.~\ref{fig:o3hb_vs_s2o3_pn}.

In Appendix \ref{section:appendix_pn} we plot the positions of the confirmed and candidate PNe on the MUSE [\ion{O}{iii}]/H$\alpha$ map. We see that spectroscopically confirmed PNe are preferentially located outside the bright \ion{H}{ii} regions (purple contours in Fig.~\ref{fig:PN}), whereas PNe candidates are distributed throughout the disc. This indicates that in bright, crowded regions the superior spatial resolution of HST data can improve the detection of PNe over low-spatial resolution 3D spectroscopy.

\par In Fig.~\ref{fig:o3hb_vs_s2o3_pn} we show the position of the confirmed PNe (in red), candidate PNe (in orange) and \hii regions (in blue) in an [\ion{O}{iii}]/\hb\ vs [\ion{S}{ii}]/[\ion{O}{iii}] diagram. We see that the spectroscopically confirmed PNe occupy a well-defined region in the diagram, as recently observed by \citet{mcleod21} in NGC~300. In contrast, candidate PNe tend to have less extreme line ratios that overlap with the \hii region portion of the diagram. However, the candidate PNe are so few in number and so low in flux levels that the contamination of the \hii region assessments can be ignored in what follows.

\par We cross match our sample of \ha\ bright regions with the spectroscopically confirmed PNe, and remove from the \hii region sample eight regions that coincide with the position of a confirmed PN along the line of sight. 

\begin{figure}
    \centering
    \includegraphics[width = \linewidth]{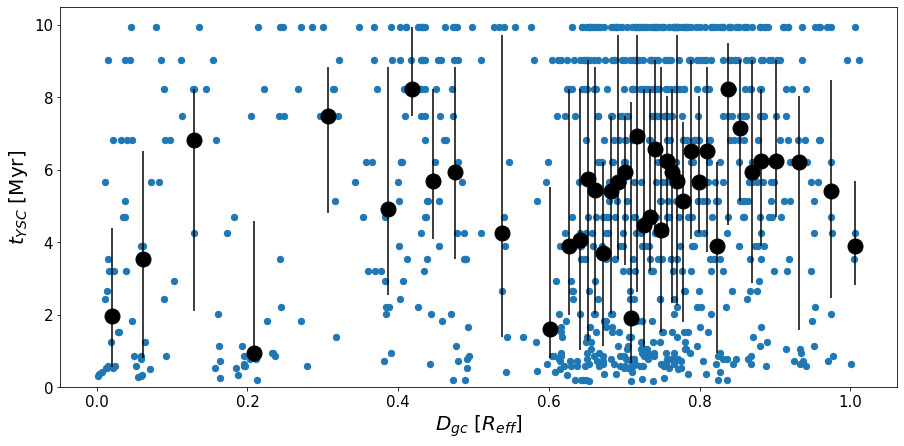}
    \includegraphics[width = \linewidth]{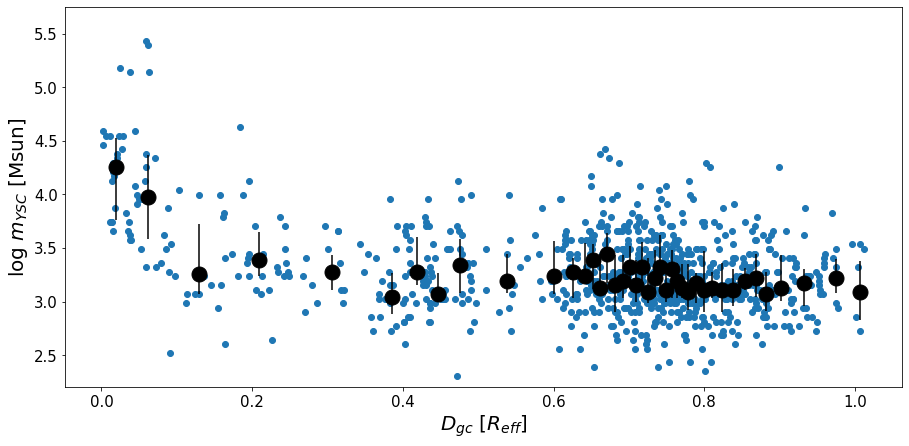}
    \caption{Age and mass of the YSCs as function of galactocentric radius. The black dots indicate the median ($\pm$ quartiles) over radial bins with equal number of objects ($\simeq$ 20).}
    \label{fig:ysc_age_mass_vs_r}
\end{figure}

\begin{figure*}
    \centering
    \includegraphics[width = \linewidth]{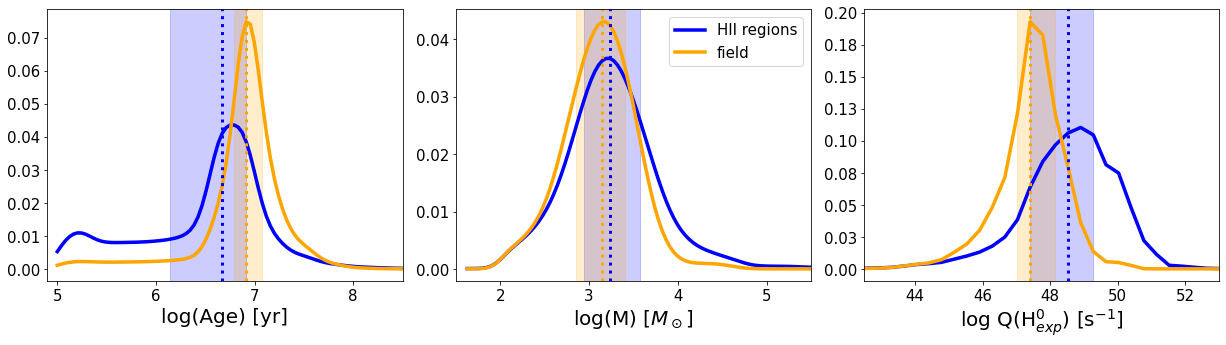}
    \caption{Combined posterior probability distributions of age, mass and ionising photon flux obtained with \textsc{SLUG} for young clusters (t $\leq$ 10 Myr) populating \hii regions (in blue) and located outside the \hii regions (`field' clusters, in orange). The vertical lines and shaded areas indicate the median and quartiles of each distribution.}
    \label{fig:combined_pdf}
\end{figure*}

\section{Stellar population in the \hii regions}
\label{section:stellar_population}

\subsection{Young star clusters}
\label{section:ysc}
Using HST narrow and broad band imaging ranging from the UV to near-infrared (NIR), \citet{silva-villa14} and \citet{adamo15} identified YSCs in the radial range 0.45 -- 4.5 kpc (0.1 -- 1.3~$R_e$). This catalogue is complete down to a few thousand of \msun\ in the age range 1 -- 10~Myr \citep[see][]{adamo15}.
In this paper, we extend the analysis of the cluster population to the inner 0.45 kpc of the galaxy, which was previously excluded due to its high luminosity gradient. We use the same dataset as in \cite{adamo15}, consisting of the F336W, F438W, F555W, F657N, and F814W WFC3 bands. Cluster candidate identification and extraction has been performed with the same software developed to analyse star cluster populations in the HiPEEC sample \citep{adamo20}.

\par In short, the extraction step is performed with the source extraction software \textsc{SExtractor} \citep{bertin96} on the reference frame (F555W). The settings are optimised to extract point-like sources in crowded regions \citep[see][]{adamo20}. We limit the extraction region to 0.47 kpc (corresponding to a radius of 500 native pixels from the centre). Aperture photometry is performed in all the bands at the positions determined in the reference frame. We use a radius of 5 pixels (0.2\arcsec) and a local sky background annulus of 6 pixels (0.25\arcsec) radius and of 2 pixel width (0.08\arcsec). We assume Vegamag as reference system, correct all the photometry for foreground galactic extinction \citep{schlafly2011}, and apply an aperture correction in all the bands, using as reference tabulated stellar encircled energy distributions. In this initial catalogue, we retain  only sources with photometric error better than 0.3 mag in F438W, F555W, F814W.

\par In total, the positions of 3133 sources are extracted. Because of the close distance of M83, we use a concentration index (CI) criterion to separate stars (PSF-like appearance) from cluster candidates (the FWHM is larger than the stellar PSF). Following \cite{adamo2017}, we estimate the CI in the reference frame (F555W) as the difference between the magnitude of the source extracted with aperture photometry of radius 1 pixels and at 3 pixels. From the distributions of the CI, we apply a CI $\ge1.2$ mag criterion to separate stars from cluster candidates. The final automatic catalogue includes only sources that have CI $\ge1.2$ mag, an absolute magnitude in the F555W band brighter than $-6$ mag, and that are also detected in the F336W with a photometric error better than 0.3 mag. This selection results in $\sim$ 300 sources.

\par Visual inspection of these sources was performed in the same way as done in the previous star cluster catalogue published by \cite{adamo15}, where `Class 1' corresponds to compact and symmetric clusters, `Class 2' to concentrated but with some degree of asymmetry systems and `Class 3' to not-cluster (stars, interlopers, artifacts in the image, etc). In total, 179 sources have been classified as class 1 and 2 in the inner region of M83. From the combined cluster catalogue covering the entire HST mosaic, we select exclusively class 1 and 2 clusters, i.e., 7459 systems. Of these, 4317 clusters are located in the MUSE FoV. We cross-match the location of the YSCs with the sample of \hii regions determined in Sect.~\ref{section:hii_regions} and find that 1251 regions host at least one YSC.

\par To include stochastic effects that arise when sampling the initial mass function (IMF) of YSCs, we used the Bayesian code \textsc{SLUG} \citep[][v2]{dasilva12,krumholz15a}. The photometric tables, containing the spectral energy distributions (SEDs) of class 1 and 2 clusters are analysed using \texttt{cluster\_slug} \citep{krumholz15a}. We compute probability distribution functions (PDFs) of clusters physical parameters based on their observed HST photometry in the five filters listed above \citep[see][]{krumholz15b}, using a library of clusters simulated with \textsc{SLUG}. We consider age $t$, mass $M$ and visual extinction $A_V$ as free parameters, and assume a flat prior in $A_V$ and in $\log t$, and a $\log(M) \sim 1/M$ prior on the mass. We use the library of mock star clusters described in~\citet{ashworth18}, a Milky Way extinction law by \citet{fitzpatrick99} and the non-rotating solar metallicity stellar population models from \citet{genevamodels12}.
As a proxy for the best value of each cluster physical properties, we use the median and quartiles of the relevant PDF, following the method tested by \citet{krumholz15b}. Naturally, reducing a full PDF to a single best value can lead to biased results, especially in the case of a non-Gaussian PDF. For a detailed study on the use of different proxies, we refer to the work of \citet{krumholz15a}.
Cluster ages, masses and ionising photon luminosity ($Q(\mbox{H}^0)$) are recovered directly from the PDFs of single clusters and will be widely used in the following analyses.

\par In Fig.~\ref{fig:ysc_age_mass_vs_r} we show the age and mass of the YSCs as function of galactocentric radius. We observe that the cluster age does not correlate with radius, whereas the central region hosts more massive clusters.  We notice that near the galactic centre, due to crowding, the low mass distribution is less complete than in the rest of the disk.
In Fig.~\ref{fig:combined_pdf} we also show the combined PDFs of young clusters (age $t \leq$ 10~Myr) located inside (in blue) and outside the \hii regions (in orange). The combined PDFs are obtained by summing the fractional probability contribution of each cluster in each logarithmic age, mass and $Q(\mbox{H}^0)$ bin and re-normalising the overall distribution.
We observe that the age PDF is double peaked, due to the degeneracy between age and extinction.
Overall, clusters populating the \hii regions are on average younger (median age of 4.7 vs 8.23 Myr) and slightly more massive (median mass of $1.7$ vs $1.4 \times 10^3$~\msun), and they emit an order of magnitude more ionising photons (median $\log Q(\mbox{H}^0)$ of 48.5 vs 47.4 s$^{-1}$).

\par
\begin{figure*}
    \centering
    \includegraphics[width=19cm]{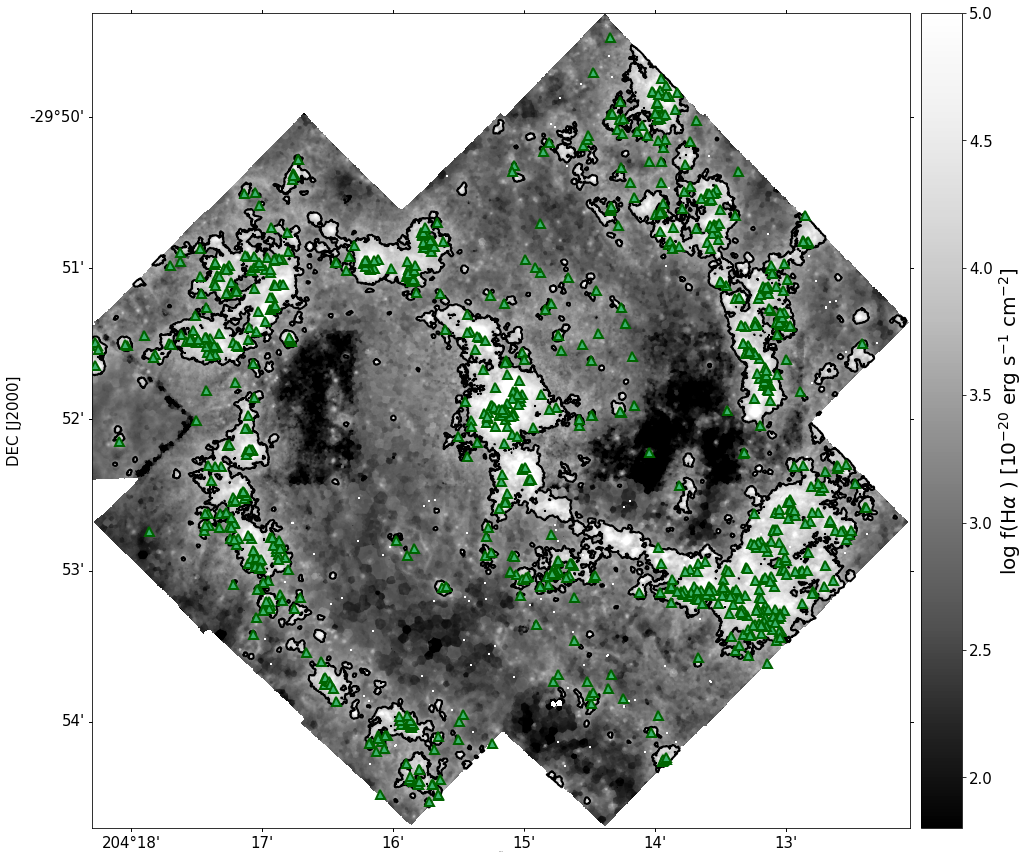}\\
    \includegraphics[width=0.49\linewidth]{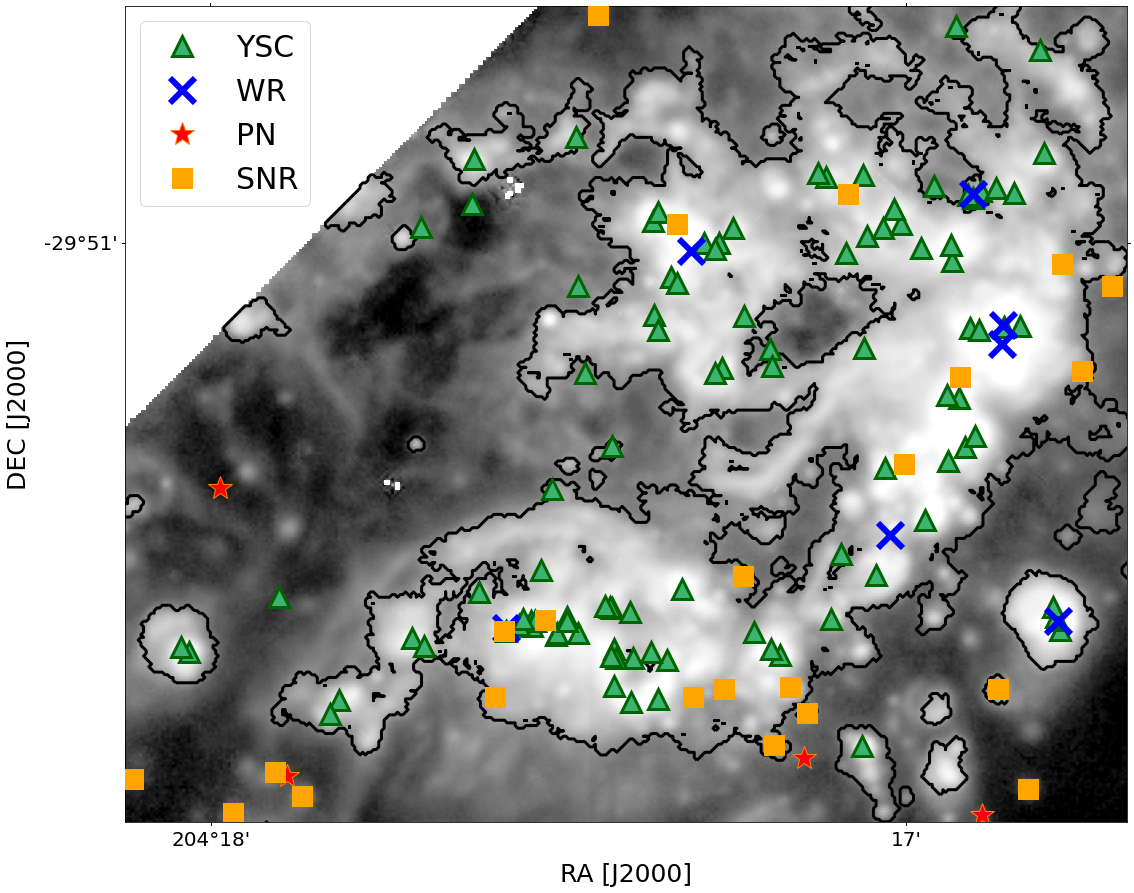}
    \caption{\textit{Top panel}: extinction-corrected H$\alpha$ map with the location of YSCs of age $\leq$ 10~Myr (green triangles). \textit{Bottom panel}: enlargement of one of the \hii region complexes. Green triangles indicate the position of YSCs observed with HST \citep{adamo15}, blue crosses and red stars indicate respectively WR stars and PNe identified in the MUSE dataset and orange squares are SNR from \citet{long2022}.}
    \label{fig:stellar_pop}
\end{figure*}

 For the remaining analysis\footnote{With the exception of the ionisation budget analysis in Sect.~\ref{section:budget}, where we sample the PDFs of all the clusters in a region, in order to not exclude YSCs with a biased median estimate. Clusters which effectively have an age $t >$ 10 Myr will not contribute significantly to the total budget.}, we only consider YSC of age $\leq$ 10~Myr populating the \hii regions, as older clusters are generally not associated with \hii regions and are more likely line-of-sight objects. We find 885 clusters of age $t \leq$ 10~Myr in the MUSE FoV, populating 532 \hii regions. Most regions host one or two clusters, but we observe up to six clusters per region in a few cases. We also find that $\sim$ 10\% of the YSCs with $t \leq 10$~Myr are located outside an \hii region. These clusters have on average a low mass ($\sim$ 1000~\msun), indicating that they have a low probability of forming massive stars. The cluster population across the MUSE FoV is shown in the top panel of Fig.~\ref{fig:stellar_pop}; we observe that the clusters trace quite closely the location of the \hii regions.

\subsection{WR stars}
\label{section:wr}
\citet{hadfield05} identified WR stars in M83 by observing photometrically selected candidates using multi-object spectroscopy. We confirm and add to this sample using spectroscopic information from the MUSE data.
We identify candidate WR stars in a \ion{He}{ii}~$\lambda$4686 linemap, obtained by integrating the gas cube over the (restframe) wavelength range 4686 -- 4695~\AA{}.
We select sources using \textsc{astrodendro}, with a flux threshold of 7.85 $\times 10^{-19}$ erg s$^{-1}$ cm$^{-2}$ (10$\sigma$ above the mean background across the map), and with a minimum leaf area of $5 \times 5$ spaxels.
We also add to this sample candidates from \citet{hadfield05} that we  missed.
For each of the 457 candidates, we extract spectra with a circular aperture of 1\arcsec\ on the full datacube (without continuum subtraction). We visually inspect the spectra, looking for characteristic features: the blue bump (BB) of \ion{He}{ii}~$\lambda$4686, \ion{C}{iii}/\ion{C}{iv}~$\lambda$4650/4658 and the red bump (RB) of \ion{C}{iv}~$\lambda$5801,5812.
We also determine if other characteristic lines, such as \ion{C}{iii}~$\lambda$5696, \ion{N}{V}~$\lambda$4603--20 and \ion{N}{iii}~$\lambda$4634--41 are present.
We confirm 68 candidates, of which 27 already identified by \citet{hadfield05}.
We further classify the spectra into helium-dominated WN type and carbon-dominated WC type, based on their spectral features.
Like \citet{hadfield05}, we find that most of the confirmed WR are late WC stars with \ion{C}{iii}~$\lambda$5696 present in their spectra.
The position and coordinates of the confirmed WR and their spectral classification can be found in appendix~\ref{section:appendix_wr} (Fig.~\ref{fig:WR_cc} and Table~\ref{table:WR_cc}).
We find that 64 of the confirmed WR are located in an \hii region. As already reported in \citet{dellabruna21} for one WR, we postulate that the remaining 4 WRs, which do not coincide with the position of \hii regions along the line of sight, are probably runaway stars. 

\bigskip In the bottom panel of Fig.~\ref{fig:stellar_pop} we show an overview of the stellar population of a single \hii region complex. We indicate the position of YSCs ($t \leq 10$~Myr, green triangles), Wolf-Rayet stars (WR, blue crosses), SNR (yellow squares) and PNe (red stars).

\section{Physical Properties of the \hii regions}
\label{section:hii_properties}

\begin{table}
\caption{Overview of the \hii regions sample. We outline the number of regions which contain at least a YSC or a WR star. This subsample is used in the analyses presented in Sections~\ref{section:P_vs_ysc} and \ref{section:budget}.}
\begin{tabular}{ll}
\hline \hline
Tot. \# of \hii regions & 4679\tablefootmark{(a)} \\
 Regions hosting YSCs with $t \leq 10$~Myr & 531 \\
 Regions hosting WRs (of which without YSC) & 63 (10) \\[0.05cm]
Tot. \# regions hosting YSC and/or WR & 541\tablefootmark{(b)} \\
\hline
\end{tabular}
\tablefoot{
\tablefoottext{a}{Of which 4654 have valid $n_e$ measurements and 149 overlap with SNRs.}
\tablefoottext{b}{Of which 499 have valid $n_e$ measurements and 42 overlap with evolved SNRs.}
}
\label{table:hii_sample}
\end{table}

In Table~\ref{table:hii_sample}, we summarise our \hii region sample.  After removing for PNe contaminants the final catalogue contains a total number of 4679 regions. Of these regions, 531 host at least 1 YSC of age $\leq$ 10~Myr, and 64 host a WR star (of which 10 regions do not contain a YSC). The fact that many regions do not host YSCs might be partly due to a bias in detecting low mass clusters, as well as the fact that many star-forming regions host less compact clustered star formation  (e.g. OB associations). In general, the cluster formation efficiency varies between 30 and 8 \% from the centre to the disk of M83 \citep{adamo15} and overall it can explain why only about $\sim10$\% of the regions host compact young clusters.

\begin{figure}
    \centering
    \includegraphics[width=0.6\linewidth]{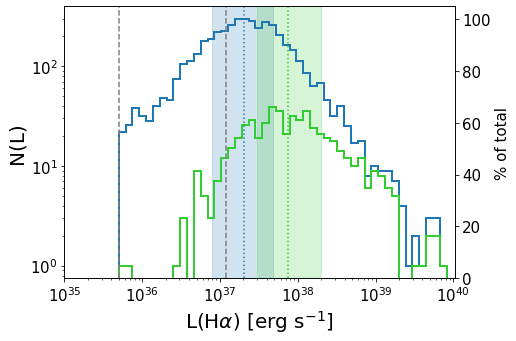}
    \includegraphics[width=0.6\linewidth]{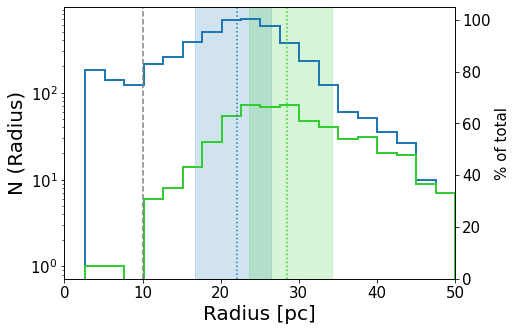}
    \caption{Luminosity function (top) and size distribution (bottom) of the final sample of \hii regions. In blue we show the total distribution, and in green the distribution for regions hosting either a YSC younger than 10~Myr or a WR star.
    The vertical lines and shaded areas indicate the median and quartiles of each distribution. The grey dashed lines in the top panel indicate the minimum luminosity of a region of radius 10 and 50 pc based on the SB threshold used for the \hii region selection. In the bottom panel, the grey dashed line corresponds to the spatial resolution limit of the MUSE data (regions below this line are unresolved in our dataset).}
    \label{fig:hii_size_lum_distr}
\end{figure}

\begin{figure}
    \centering
    \includegraphics[width=\linewidth]{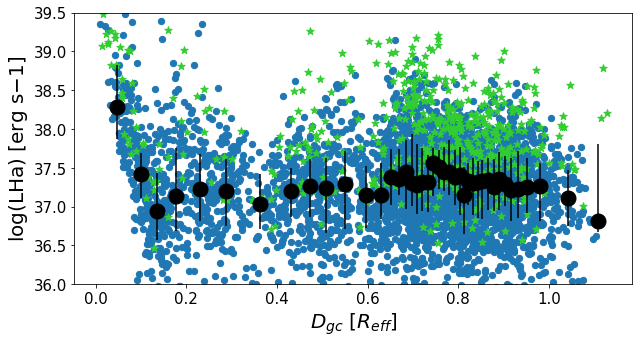}
    \includegraphics[width=0.92\linewidth]{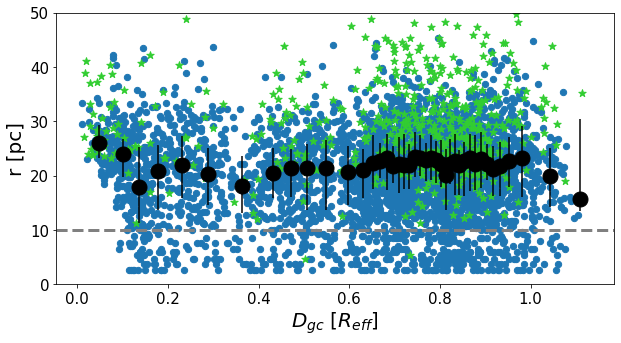}
    \caption{Luminosity (top panel) and size (bottom panel) of the final sample of \hii regions as function of galactocentric radius. Green stars indicate regions hosting either a YSC younger than 10~Myr or a WR star. The large black dots indicate the median ($\pm$ quartiles) in radial bins with equal number of objects ($\simeq$ 120). The dashed line in the bottom plot indicates the spatial resolution limit of the MUSE data.
    }
    \label{fig:hii_L_size}
\end{figure}

\begin{figure}
    \centering
    \includegraphics[width=\linewidth]{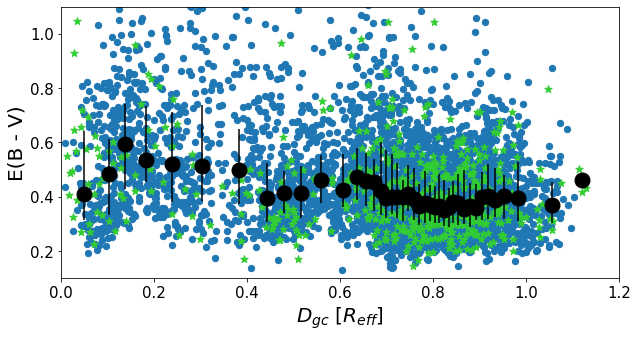}
    \includegraphics[width=\linewidth]{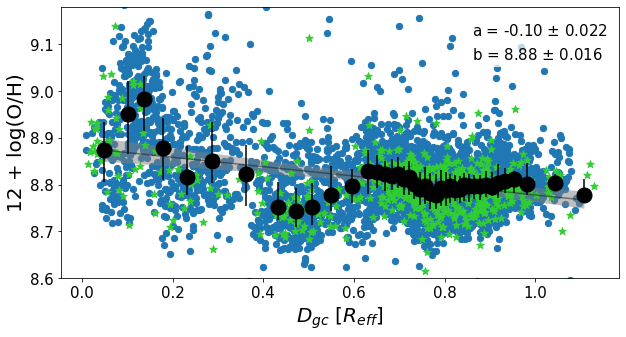}
    \includegraphics[width=\linewidth]{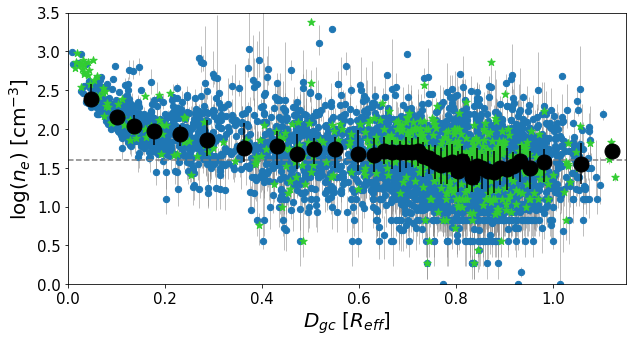}
    \caption{Same as Fig.~\ref{fig:hii_L_size} but showing the extinction (top), oxygen abundance (centre) and density (bottom panel) of the \hii regions. In the central panel, we show a linear fit to the data (in black) with a 95\% confidence interval (shaded grey area). We also indicate the best fit parameters in the upper right corner. In the bottom panel, the grey dashed line in the bottom panel indicates the sensitivity limit of the density determination method: points below this lines are consistent -- within measurement uncertainty -- with $n_e \simeq$ 1 cm$^{-3}$. Error bars span the first to third quartile of a distribution of $n_e$ obtained from 1000 Monte Carlo realisations of the [\ion{S}{ii}] ratio (see Sect.~\ref{section:hii_properties}).
    }
    \label{fig:hii_ebv_z_ne}
\end{figure}

\begin{figure}
    \centering
    \includegraphics[width=\linewidth]{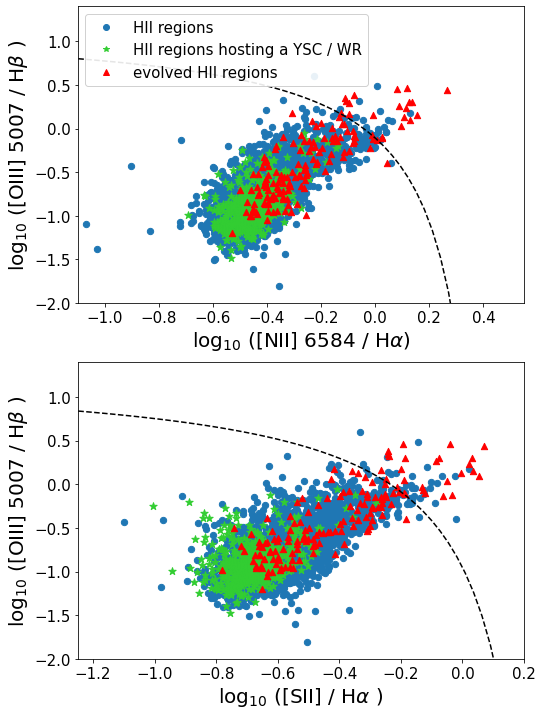}
    \caption{Location of the final sample of \hii regions in BPT emission line diagrams. Green stars indicate the subset of regions hosting either a YSC younger than 10~Myr or a WR star. Red triangles indicate evolved regions hosting a SNR. The black dashed line indicate the `extreme starburst line' from \citet{kewley01}, denoting the upper limit for gas purely excited by star-formation.}
    \label{fig:hii_bpt}
\end{figure}

In total, we have access to the stellar population of 541 regions (42 of which are evolved regions hosting a SNR\footnote{We remark that in 18 of these regions we observe an \ha\ velocity dispersion $> 40$~km s$^{-1}$ ($> 60$~km s$^{-1}$ for 5 of the regions), indicative of the fact that shocks might be playing a non-negligible contribution to the \ha\ emission.}). This is the subset of regions used in most of the remaining analysis, whose properties are highlighted in light green in Fig.~\ref{fig:hii_size_lum_distr} -- \ref{fig:hii_ebv_z_ne} and Fig.~\ref{fig:hii_bpt}.

\par In Fig.~\ref{fig:hii_size_lum_distr} we show the luminosity function (top panel) and size distribution (bottom panel) of the regions. The radius of each \hii region is estimated by approximating the area enclosed by the boundaries determined in Sect.~\ref{section:hii_identification} to a circular area. We observe that regions with YSCs (in light green) are on average more luminous and larger in size. The recovered luminosity distribution is comparable to the range reported by \cite{kreckel19} and \citet{santoro22} for the PHANGS-MUSE sample. The `turnover' at $L_{min} \sim 2 \times 10^{37}$ erg s$^{-1}$ is due to incompleteness \citep[see][]{kennicutt89} and is consistent with the values of $L_{min}$ obtained by \citet{santoro22} in nine of the PHANGS galaxies. The incompleteness originates on one hand from the non-detection of faint regions, due to both instrumental sensitivity limit and the SB threshold used in the \hii regions detection. On the other hand, there are blending effects of low-luminosity regions with neighboring bright objects, especially in crowded and luminous region complexes, for example near the galactic centre.
The distribution of radii peaks at much smaller sizes than in PHANGS-MUSE \citep{kreckel19}, as a result of the higher spatial resolution of our dataset ($\sim$ 20 pc) compared to the average resolution of PHANGS-MUSE ($\sim$ 50 pc). The distribution of radii observed in this work is comparable with sizes reported in other local galaxies at resolution $<$ 20 pc \citep[e.g.][]{sivan90, mcleod21}. Finally, we assess that (for $r > 20$ pc) the frequency distribution of diameters is well fitted by a \citet{vandenbergh81} exponential law
$$N = N_0 \exp{-D/D_0},$$
with $N_0 \sim 8$ and $D_0 \sim 45$ pc. This is in agreement with observations in most spiral galaxies \citep{ye92, gutierrez11, azimlu11, souza18}.

\par In Fig.~\ref{fig:hii_L_size}, we show how the size and luminosity of the regions vary as function of galactocentric radius.
The most luminous regions are located within the inner 0.5 kpc ($R \leq 0.15~R_e$) of the galaxy, coincident with the starburst region (see \paperI). A noticeable increase in \hii region luminosity is also observed in correspondence of the highly star-forming regions located at the end of the bar at $\sim 2.3$~kpc (0.7~$R_e$), the same region where the cluster formation efficiency is observed to increase as reported in \cite{adamo15}. We observe that the radius of the regions does not change significantly as a function of galactocentric distance.

\begin{figure}
    \centering
    \includegraphics[width=\linewidth]{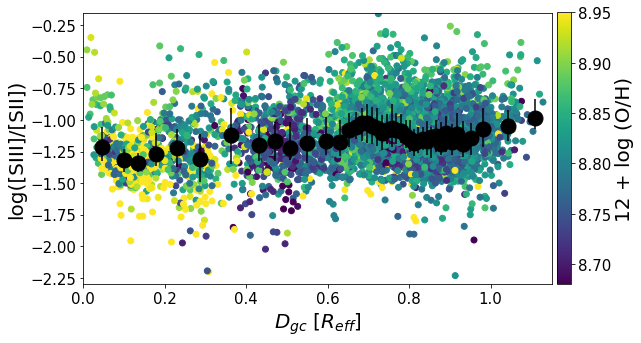}
    \caption{Ratio of [\ion{S}{iii}]~$\lambda$9069/[\ion{S}{ii}]~$\lambda\lambda$6716,31 (reddening corrected) of each \hii region as function of galactocentric radius. This ratio is a tracer of the hardness of the radiation field within the \hii regions. }
    \label{fig:hii_s3s2}
\end{figure}

\par Fig.~\ref{fig:hii_ebv_z_ne} shows the average extinction, oxygen abundance and electron density of the \hii regions as function of galactocentric radius. In order to compute these quantities, we first obtained an integrated spectrum for each region. We then fit single Gaussian profiles to the emission lines (see~\paperI\ for details). In the case of the [\ion{S}{iii}]~$\lambda$9069 line, which was not in the wavelength range for which we had removed the stellar continuum (see \paperI), we manually subtracted a median stellar continuum determined in the (restframe) wavelength range 9034 -- 9054~\AA{} and 9079 -- 9099~\AA{}.
\par We determine the extinction (top panel in Fig.~\ref{fig:hii_ebv_z_ne}) from the \ha/\hb\ ratio using \textsc{pyneb} \citep{luridiana15}, assuming an intrinsic ratio of 2.863 (case B recombination with $T_e = 10^4$~K, $n_e = 100$~cm$^{-3}$, \citealp{osterbrock}) and a \citet{cardelli89} extinction law. We see that the average extinction peaks within the circumnuclear starburst region (E($B - V$) $\sim$ 0.6) and decreases slightly with radius, down to $\sim$ 0.4 mag.
\par The oxygen abundance (central panel in Fig.~\ref{fig:hii_ebv_z_ne}) was determined from the [\ion{N}{ii}]~$\lambda$6584/\ha\ := N2 ratio, using the calibration from \citet{denicolo02}:
$$ 12 + \mbox{log(O/H)} = 9. +  0.73 \times \mbox{N2}.$$
We decided to use the N2 ratio for two reasons. First of all, \citet{bresolin16} studied the metallicity of \hii regions in M83 using different line ratios and calibrations, and found that the N2 ratio with the calibration above gives an excellent agreement with the metallicity recovered using direct methods and determined from observations of blue supergiants.
Secondly, we wanted to avoid using multiple lines which would result in stronger degeneracies with temperature and density \cite[see e.g.][]{ercolano12, mcleod16a, mcleod19, dellabruna21}.
The recovered range in oxygen abundance agrees well with the results of \citet{bresolin16} for the radial range probed by the MUSE data (corresponding roughly to $0.4~R_{25}$ in their work). We perform a linear fit to the data and obtain
$$12 + \mbox{log(O/H)} = 8.88 (\pm 0.016) - 0.10 (\pm 0.033)~R/R_e,$$
in very good agreement with the central abundance of 8.87 and slope of $-0.09$ reported by \citet{bresolin16} for the radial range 0 -- 3~$R_{25}$ (0 -- 8~$R_e$) using the N2 ratio. We note large variations as a function of galactocentric distance. The abundance peaks at $12 + \mbox{log(O/H)} \sim 9.0$ at about 0.14 R$_{\rm eff}$, declines to an average value of $12 + \mbox{log(O/H)} \sim 8.7$ in the interarm region (0.3 -- 0.6 R$_{\rm eff}$) and plateaus around 8.8 at larger radii, which are dominated by the presence of the spiral arms. Further analysis of the metallicity distribution as a function of spatial position will be presented in an upcoming work (Adamo et al., in prep.).

\par Finally, the electron density (bottom panel in Fig.~\ref{fig:hii_ebv_z_ne}) is determined from the ratio of [\ion{S}{ii}]~$\lambda\lambda$6716,6731, using \textsc{pyneb}. We remark that densities below $\lesssim$ 40 cm$^{-3}$ (grey dashed line in the figure) are in the asymptotic part of the [\ion{S}{ii}] vs $n_e$ curve and are consistent -- within the uncertainties -- with the lowest density limit probed by the ratio, $n_e \sim 1$ cm$^{-3}$ \citep[see e.g.][]{kewley19}. To take into account uncertainties on the method, we considered 1000 Monte Carlo realisations of the [\ion{S}{ii}] ratio within the measurement errors, adopt as fiducial value the median of the resulting distribution and indicate the first to third quartile as grey error bars.
The density of the \hii regions decreases by almost two orders of magnitude between the centre of the galaxy and the outer regions. This trend seems to tightly follow the decline in the gas surface density of the molecular gas reported in \citet{lundgren04} and \citet{adamo15}, as well as in the average midplane pressure reported in Fig.~\ref{fig:P_comparision}, suggesting that the ambient environment where \hii regions form might be responsible for some of their key physical properties \citep{smith06}.

\par In Fig.~\ref{fig:hii_s3s2} we furthermore inspect the ionisation state of the regions as function of radius. We use as tracer the ratio of [\ion{S}{iii}]/[\ion{S}{ii}]. A high ratio indicates gas with a high ionisation state, where doubly ionised sulphur is dominant, whereas a low ratio is indicative of singly ionised ions (low ionisation state).
In general, we observe that regions at larger radii have a higher ratio of [\ion{S}{iii}]/[\ion{S}{ii}] (higher ionisation state). This is expected based on the slight decrease in metallicity with radius, as shown in the colour axis, and is due to the fact that stars at lower metallicity are hotter and therefore emit harder ionising radiation.

\par Finally, in Fig.~\ref{fig:hii_bpt} we place the \hii regions on N2- and S2-`BPT' emission line diagrams \citep{baldwin81, veilleux87}. As described in \citet{kewley19_review}, the position occupied by an object in these diagrams is a function of many different parameters, such as the density, radiation strength and hardness and metallicity. These diagrams were originally devised to identify the source of ionisation in single aperture spectra of galaxies but are today also used to determine the source of ionisation in different emission line regions within galaxies. \citet{kewley01} determined using models of star forming galaxies and shocks what is the upper limit spanned by purely star-forming galaxies in the diagrams (`extreme starburst line'). In Fig.~\ref{fig:hii_bpt} we confirm that all the regions lie below or very close to this limit. We observe that most regions beyond the extreme starburst line are evolved \hii regions hosting one or more SNR (red triangles), and that regions hosting young clusters or WR stars (green dots) are all located below this threshold.

\section{Pressure analysis}
\label{section:pressure}
In order to investigate which feedback mechanisms are dominant in each region, we study the contribution of the two main pre-SN feedback mechanisms (ionised gas pressure and radiation pressure) to the region's internal pressure.
We then compare how these vary with radial position in the galaxy and with the properties of the YSCs powering the regions.

\begin{figure*}
    \centering
    \includegraphics[width = 0.7\linewidth]{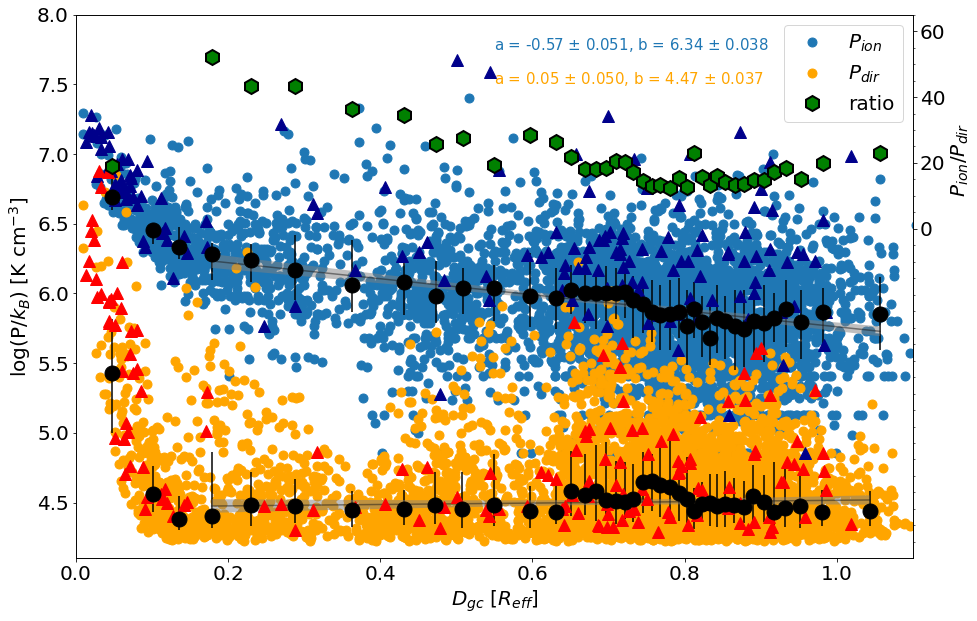}
    \caption{Ionised gas pressure (in blue), direct radiation pressure (in orange) and their ratio (in green, values indicated on the right axis) as function of galactocentric radius. Red and dark blue triangles indicate evolved \hii regions (hosting a SNR). The black dots indicate the median ($\pm$ first and third quartile) pressure in radial bins containing an equal number of points ($\sim$ 120). On top of the figure, we indicate the coefficients of a linear fit to the data.}
    \label{fig:P_vs_r}
\end{figure*}

\subsection{Ionised gas pressure}
\label{section:Pion}
The pressure exerted by the warm ionised gas is simply described by the ideal gas law:
$$P_{\rm ion} = (n_e + n_H + n_{He}) k_B T_{\rm HII},$$
where $T_{\rm HII}$ is the temperature of the \hii gas. Assuming singly ionised Helium, this simplifies into $P_{\rm ion} \approx 2 n_e k_B T_e$. We computed the electron density using \textsc{pyneb} and the ratio of [\ion{S}{ii}]~$\lambda$6716,6731. Given that in the MUSE dataset we do not detect any temperature sensitive line with sufficient S/N, we assume a constant temperature $T_e = 10^4$~K.

\subsection{Direct radiation pressure}
\label{section:Pdir}
Direct radiation pressure is the pressure exerted by the momentum of the photons in the region. The volume-averaged direct radiation pressure can be derived from the observed total (bolometric) reddening-corrected luminosity $L_{\rm bol}$ as described by \citet{lopez14}:
$$P_{\rm dir} = \frac{1}{V} \int P_{rad} dV = \frac{3 L _{bol}}{4 \pi R^2 c},$$
where $R$ is the radius of the region. We note that this is an upper limit to the effective radiation pressure in the region, assuming a `classic' Str\"omgren sphere morphology with an optically thick envelope (ionisation bounded region).
To determine $L_{bol}$ we use the relation between the bolometric luminosity and the \ha\ luminosity, $L_{bol} = 138~L(H\alpha)$, derived by \cite{kennicutt_evans12}, under the assumption of a stellar population with a fully sampled\footnote{We remark that this might not be the case for the fainter \hii regions in our sample. This could result in a slight overestimation of $L_{\rm bol}$ and hence of $P_{\rm dir}$, which does however not significantly affect our results (given the several orders of magnitude spanned by $P_{\rm dir}$).} IMF in the age range 0 -- 10 Myr.

\subsection{Trends with galactic radius}
\label{section:P_vs_r}

\begin{figure*}
    \centering
    \includegraphics[width = 0.49\linewidth]{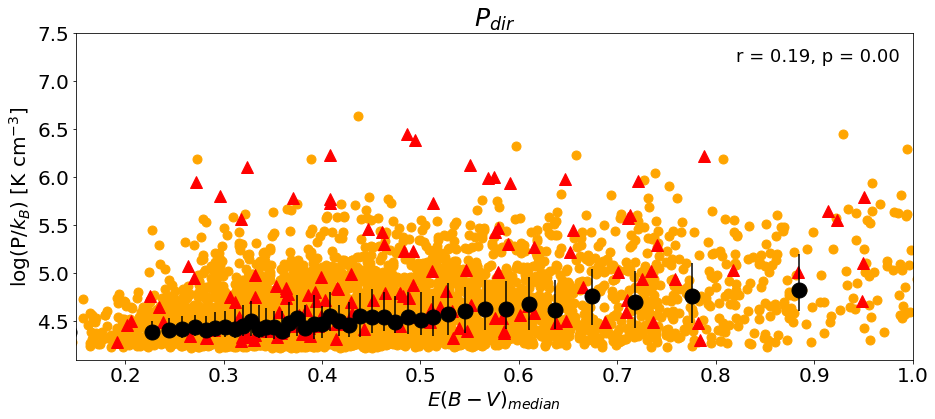}
    \includegraphics[width = 0.49\linewidth]{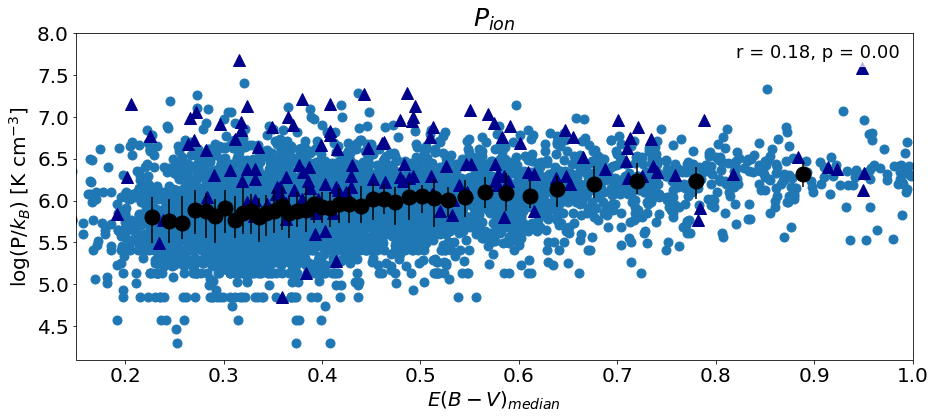}\\
    \includegraphics[width = 0.49\linewidth]{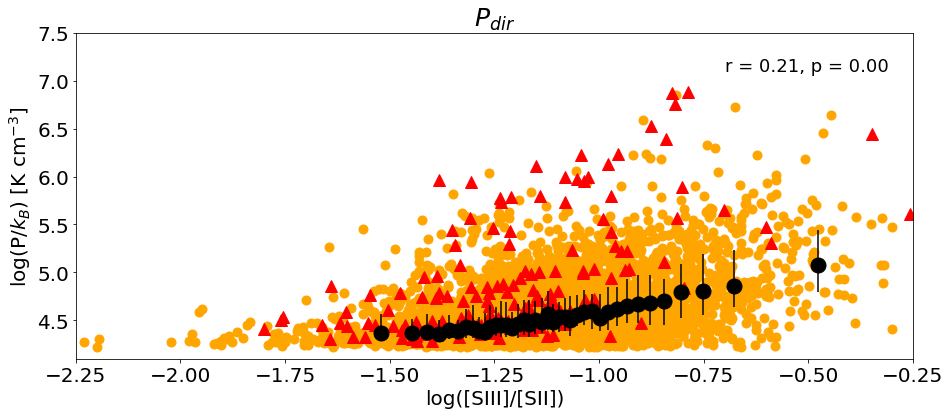}
    \includegraphics[width = 0.49\linewidth]{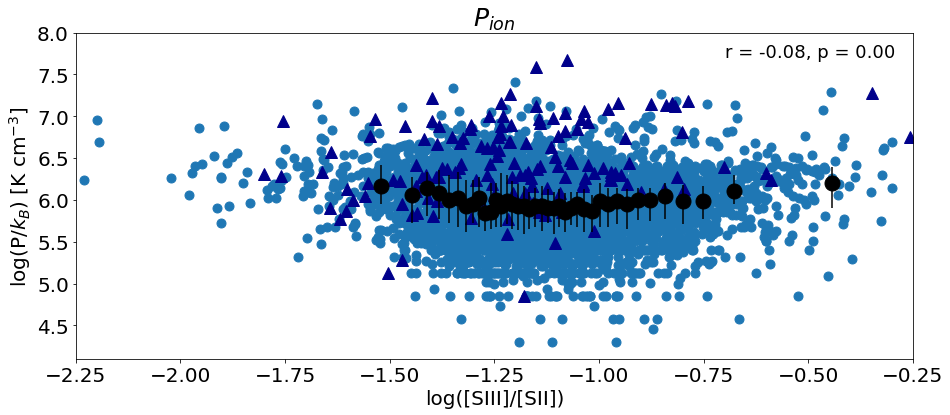}
    \caption{Pressure terms as function of average reddening (top panels) and radiation field hardness (bottom panels) in the regions. Red and blue triangles indicate evolved \hii regions (hosting a SNR). The black dots indicate the median ($\pm$ quartiles) pressure in radial bins containing an equal number of points. On the top right, we indicate the value of the Pearson's correlation coefficient $r$.}
    \label{fig:P_vs_ebv_Z}
\end{figure*}

In this work we focus on global environmental dependencies described by trends as a function of 1D galactocentric distance. In a follow-up paper we will investigate the impact of spiral arms by performing a 2D analysis which retains the azimuthal information about arm and interarm environments.
\par In Fig.~\ref{fig:P_vs_r}, we show $P_{\rm ion}$ (in blue), $P_{\rm dir}$ (in orange) and their ratio (in green) as function of galactocentric radius.
In general, $P_{\rm ion}$ dominates over $P_{\rm dir}$, with a median ratio $\simeq$ 13 across our sample. This is in agreement with observations of nearby galaxies \citep{lopez11, lopez14, mcleod19, mcleod20, mcleod21, barnes21} at scales of 10 -- 100s parsec.
Environmental dependencies are also observed. Both pressure terms are enhanced in regions located within the central starburst ($D_{\rm gc} \lesssim 0.15~R_e$). In particular, P$_{\rm dir}$ is up to to 2 order of magnitude higher with respect to \hii regions located in the disk, becoming comparable to $P_{\rm ion}$. An increase in P$_{\rm dir}$ is also observed in the end-of-bar region at $\sim0.7~$R$_{\rm eff}$. Outside the central starburst, we observe that $P_{\rm ion}$ is decreasing as a function of radius (linear fit slope $a \simeq -0.6$, Pearson's correlation coefficient $r = - 0.48$), whereas $P_{\rm dir}$ stays approximately constant ($a \simeq 0.0$, $r = -0.15$).

\par In Fig.~\ref{fig:hii_ebv_z_ne} and~\ref{fig:hii_s3s2} we see that -- despite strong local variations -- overall the extinction is decreasing and the radiation field hardness is increasing with galactocentric radius. In order to investigate whether radial trends observed in the pressure terms could be related to the radial variation in these quantities, in Fig.~\ref{fig:P_vs_ebv_Z} we plot the pressure against the average reddening and radiation field hardness of each \hii region.
\par For $P_{\rm dir}$ we observe both a positive correlation with reddening ($r = 0.19$) and with radiation field hardness ($r = 0.21$). Similar trends have also been reported by \cite{mcleod21} for the flocculent galaxy NGC 300. The positive correlation between $P_{\rm dir}$ and reddening is produced by a positive correlation between $L_{\rm bol}$ and E($B - V$). In general, we find that luminous \hii regions have higher reddening and harbour a harder radiation field, resulting therefore in higher $P_{\rm dir}$ exerted within the \hii regions.

\par On the other hand, the trends with $P_{\rm ion}$ require a more complex interpretation. We see a positive correlation with extinction ($r = 0.18$) and a weak anti-correlation with radiation field hardness ($r = -0.08$).
In this case, the correlation with extinction can be interpreted in the light of the fact that regions with higher values of E$(B - V)$ are typically at a younger evolutionary stage (see also Sect.~\ref{section:P_vs_ysc}) and are hence more compact, resulting in a higher $n_e$ and therefore an increase in $P_{\rm ion}$ (at constant temperature). 
The results of radiation-hydrodynamical models from \citet{ali21} discussed in \citet{mcleod21} agree with this scenario, indicating that an increase in UV photon extinction results in smaller regions.
We directly investigate the correlation between \hii region radius and $P_{\rm ion}$, finding a modest negative correlation ($r = -0.13$). The trend is probably weakened by the presence of both very young and compact regions as well as the smaller \hii regions surrounding older stars. 
The anti-correlation with radiation field hardness can be understood in the light of two factors. First, a harder radiation field is caused by a lower metallicity (Fig.~\ref{fig:hii_s3s2}), which in turn results in higher electron temperatures (whereas we have assumed a constant $T_e$ in the computation of $P_{\rm ion}$). Taking this effect into consideration will weaken the observed correlation. Second, the regions with the highest metallicity are on average located closer to the centre of the galaxy (see central panel Fig.~\ref{fig:hii_ebv_z_ne}), where $n_e$ is also enhanced (bottom panel Fig.~\ref{fig:hii_ebv_z_ne}). Thus, the observed trend between $P_{\rm ion}$ and [\ion{S}{iii}]/[\ion{S}{ii}] is likely in large part driven by the dramatic impact that the galactic environment has on the physical properties of rapidly evolving star-forming regions and only to a second degree to variations in the intrinsic properties of the regions. We discuss this further in Sect.~\ref{section:discussion}.

\begin{figure*}
    \centering
    \includegraphics[width = 0.49\linewidth]{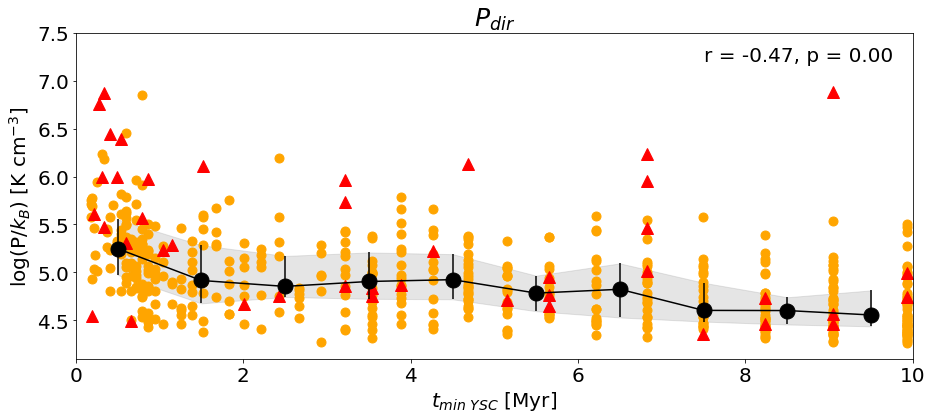}
    \includegraphics[width = 0.49\linewidth]{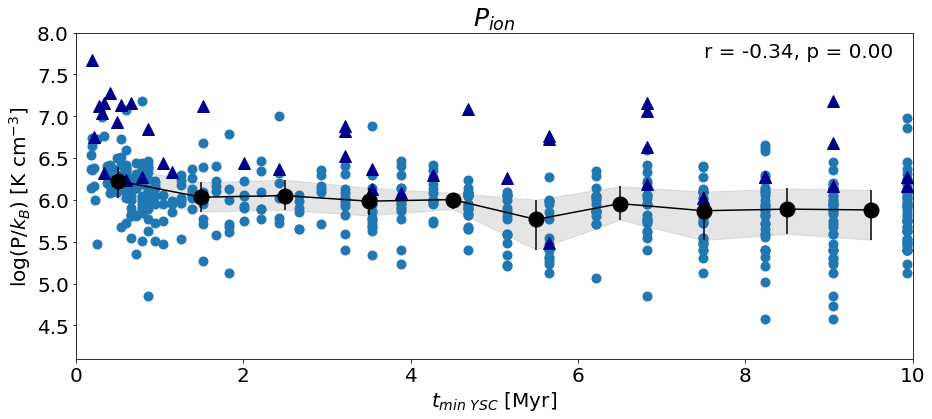}\\
    \includegraphics[width = 0.49\linewidth]{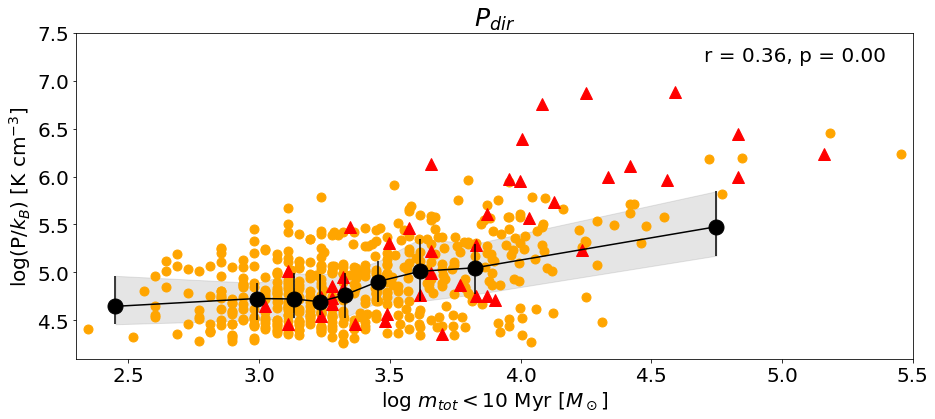}
    \includegraphics[width = 0.49\linewidth]{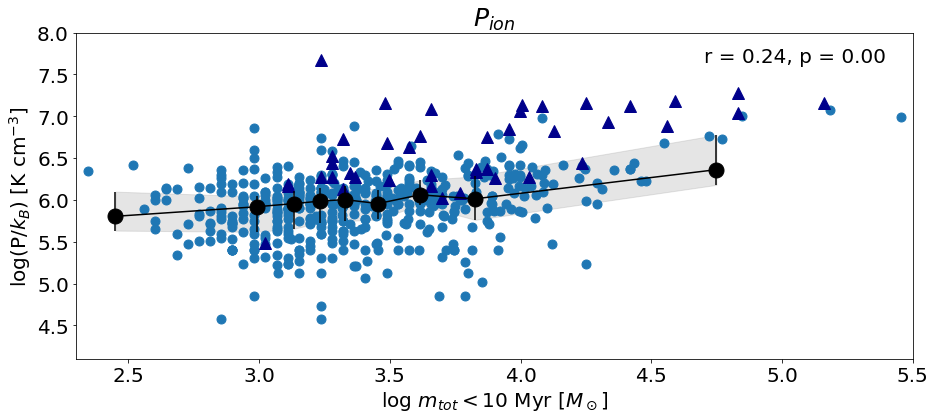}
    \caption{Pressure terms as function of the age of the youngest cluster in the region (top) and the total mass of clusters younger than 10~Myr (bottom).
    Red and blue triangles indicate evolved regions hosting a SNR. The black dots and line indicate the median ($\pm$ quartiles, shaded in grey) in radial bins of 1~Myr (top) and in mass bins with equal number of objects (bottom, $\simeq$ 40 objects/bin). On the top right, we indicate the value of the Pearson's correlation coefficient $r$.}
    \label{fig:P_vs_ysc}
\end{figure*}

\subsection{Trends with YSC properties}
\label{section:P_vs_ysc}
As observed in Fig.~\ref{fig:hii_size_lum_distr}, \hii regions containing YSCs are brighter and larger in size with respect to regions that do not host any compact detected cluster. When considering the physical properties and ionisation state of these \hii regions (Fig.~\ref{fig:hii_ebv_z_ne} and \ref{fig:hii_bpt}), those hosting YSCs do not show any significant deviations from the average properties. We investigate here to what extent the pressure terms are linked to the star cluster physical properties. Extrapolating from \cite{adamo15}, we know that cluster formation efficiency changes between 8 and 30\% from the outer disk to the centre of M83. This means that star clusters represent only a fraction of the stars forming and powering the \hii regions. However, it is important to notice, as reported in several numerical simulations \citep[e.g.][]{kim17, gentry17, fielding18, bending20}
that because of the compact configuration of the stars within star clusters we expect that stellar feedback couples more efficiently with the surrounding \hii region. 

\par Because multiple YSCs can be found within the same \hii region, we plot in Fig.~\ref{fig:P_vs_ysc} the pressure terms against the age of the youngest cluster (top panels), assuming that the cluster with the youngest age has the most significant contribution to the LyC photon production in the region. We find that regions hosting younger clusters have both a higher $P_{\rm dir}$ ($r = -0.47$) and $P_{\rm ion}$ ($r = -0.34$). The two pressure terms peak in regions containing very young clusters ($\sim$ 1~Myr), although there is some scatter, probably due to the simplifying assumption that the youngest cluster is the one producing the highest photoionisation rates. We also see that after 1 to 2~Myr, $P_{\rm ion}$ remains constant, suggesting that -- whereas cluster feedback dominates the pressure terms at very young ages -- in regions hosting more evolved clusters the ionising photon flux is maintained by the young stars surrounding the clusters. On the other hand, $P_{\rm dir}$ continues to decrease with cluster age, as expected if dust is destroyed in more evolved regions, thus reducing the coupling efficiency between dust and photons flux. Moreover, as we remarked in Sect.~\ref{section:Pdir}, $P_{\rm dir}$ is an upper limit assuming an ideal ionisation bounded region, which might not be the case for more evolved regions.
In the bottom panels of Fig.~\ref{fig:P_vs_ysc}, we plot the pressure terms as a function of the total stellar mass in clusters (of age $t \leq 10$~Myr). We observe that a higher total mass in young clusters correlates with a steady increase in the pressure terms ($r = 0.36$ and 0.24, respectively). This is true especially for $P_{\rm dir}$. We discuss these trends further in Sect.~\ref{section:discussion}.

\section{Photoionisation budget}
\label{section:budget}

In the previous section, we identified a coupling between the physical properties of the very young star clusters and the pressure exerted on their host \hii regions. In this section, we evaluate whether this coupling between star cluster physical properties and pressure terms leads also to a correlation between  cluster physical properties and escape of ionising radiations from the \hii regions.  In \cite{weilbacher18}, the authors report that the ionising radiation produced by the star cluster population in the Antennae merger system is sufficient to produce the ionising radiation observed in the emitting ISM. We assess here whether this is the case also in the spiral galaxy M83. 

\par Given that we lack a catalogue of young, massive stars outside clusters, we limit the ionisation budget analysis to regions hosting at least one YSC. For completeness, we also include in the budget the ionising radiation produced by the spectroscopically confirmed WR stars, even if they do not reside in regions containing YSCs. In the latter case, we assign the region an age of 4~Myr. In total, we perform the budget analysis on 541 regions, 531 hosting YSCs and 10 hosting exclusively WR stars, as summarised in Table~\ref{table:hii_sample}.

\par In general, the rate of hydrogen ionising photons emitted from a region, $Q(\mbox{H}^0)$, is defined as the total number of LyC photons emitted per unit time
$$Q(\mbox{H}^0) = \int_{\nu_0}^\infty \frac{L_\nu}{h\nu} d\nu.$$
We infer an `observed' $Q(\mbox{H}^0)$, hereafter $Q_{\rm obs}$, for each region from the \ha\ flux, and compare it to the `expected' $Q(\mbox{H}^0)$, hereafter $Q_{\rm exp}$, obtained by modeling the ionising photon rate produced by YSCs and WR stars within the region. We then compare the two quantities to compute an escape fraction
$$f_{\rm esc} = 1 - \frac{Q_{\rm obs}}{Q_{\rm exp}}.$$
A value of $f_{\rm esc} > 0$ indicates that the observed emission is lower than the expected emission, indicating that some radiation might have escaped the region. We caution however that other effects can mimic an $f_{\rm esc} > 0$. One of these is the absorption of LyC radiation by dust: dust can absorb the LyC photons before they can ionise the hydrogen atoms, and re-emit them at longer wavelength. The typical fraction of LyC radiation absorbed by dust has been constrained to be of the order of 30 -- 50\% in observations of nearby galaxies \citep{inoue01, hirashita03, iglesias04, salim16} as well as in simulations \citep{tacchella22}.

In the following, we summarise how we estimate $Q_{\rm obs}$ and $Q_{\rm exp}$ for each \hii region. 

\subsection{Observed ionising photon flux}
We compute the observed ionising photon flux from the (dereddened) \ha\ luminosity. We use the following conversion factor from \citet{draine}, which assumes case B recombination with an electron temperature of $T_e \sim 10^4$ K and a density of $n_e = 10^3$ cm$^{-3}$:
$$ Q(\mbox{H}^0)_{\rm obs}~\mbox[s^{-1}] = 7.31 \times 10^{11} L(H\alpha) \quad [\mbox{erg s}^{-1}].$$

\subsection{Modeling YSC ionising radiation output}
We model the emission of YSCs with the \textsc{slug} Bayesian stellar population synthesis code based on their observed HST photometry, as described in Sect.~\ref{section:ysc}. 
We obtain the total ionising flux of a region by constraining a PDF of $Q(\mbox{H}^0)$ for each cluster, sampling one value from each one of these PDFs and repeating this procedure during 1000 Monte Carlo realisations, to obtain a distribution of the total flux. We then report the median and quartiles of the total distribution as best value and related uncertainty.

\subsection{Modeling WR stars}
WR stars emit very powerful radiation, and a single WR star can substantially contribute to the ionisation budget of a region \citep[see e.g.][]{dellabruna21}.
We constrain $Q_{\rm exp}$ for each star using values tabulated by \citet[][Tables 3 and 4 in their work]{smith02}.
We use the WR spectral types listed in Table~\ref{table:WR_cc} of the Appendix~\ref{section:appendix_wr} and assume a temperature range $T = [40 - 80] \times 10^3$ K for WN-type stars and $T = [100 - 140] \times 10^3$ K for WC-type stars. We obtain a distribution of $Q(\mbox{H}^0)$ from 1000 Monte Carlo realisations of $T$ and we consider its median and quartiles as best value and associated uncertainty.

\subsection{Resulting budget}

\begin{figure}
    \centering
    \includegraphics[width=\linewidth]{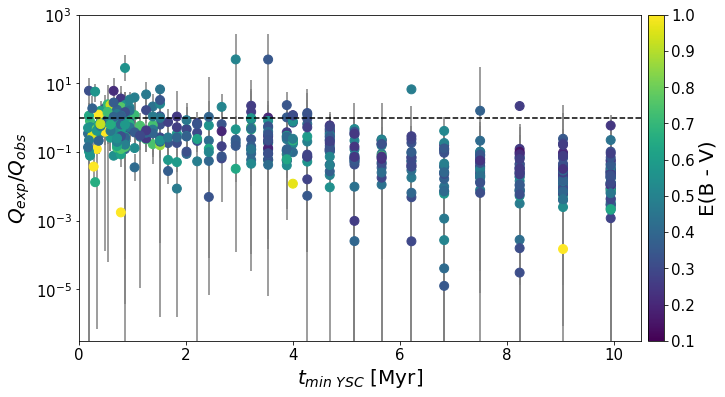}
    \includegraphics[width=\linewidth]{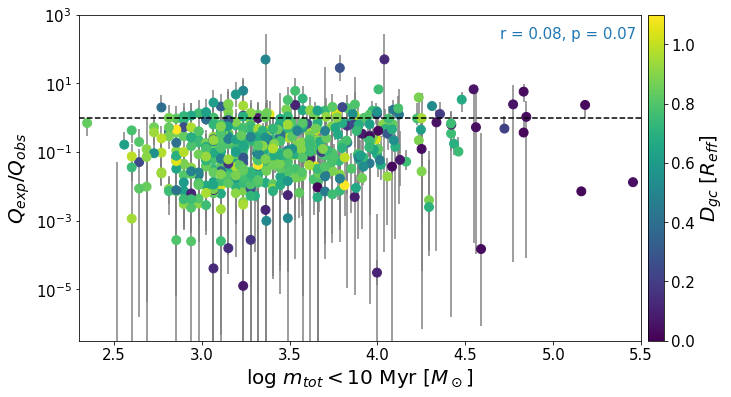}
    \caption{Ratio of expected to observed $Q(\mbox{H}^0)$ for all the \hii regions hosting YSCs or WR stars. The black dashed line indicates a ratio of 1 ($f_{\rm esc} = 0$). \textit{Top panel}: $Q$ ratio as function of the age of the youngest cluster ($t_{min}$) hosted in each region , colour coded by the region's average extinction.
    \textit{Bottom panel}: $Q$ ratio as function of the total mass in clusters younger than 10~Myr, colour coded by galactocentric distance.}
    \label{fig:qratio_vs_age}
\end{figure}

Figure~\ref{fig:qratio_vs_age} shows the resulting budget for all the regions analysed. Points above the black dashed line indicate regions with an $f_{\rm esc} > 0$. Overall, we observe escape only for $\sim$ 13\% of the regions (69 out of 541), pointing to the fact that only a fraction of YSCs does  produce enough ionising radiations to match the observed \ha\ luminosity of \hii regions. In the majority of the leaking regions, i.e. 80 \%, the youngest cluster has an age $t_{\rm min}< 2$ Myr (top panel). 
In order to quantify the effect of dust absorption, we color code each region by its average E($B - V$). We do not observe an excess in extinction among the regions with $f_{\rm esc} >0$, indicating that dust is likely not playing a dominant role, and that these escape fractions are significant.

\par We also investigate whether regions that have been forming stars for a longer time are advantaged in leaking ionising photons, motivated by the recent results of high resolution simulations from \citet[][see Sect.~\ref{section:introduction}]{ma20}. We assume that the age of the oldest star cluster in the \hii region is an indicator of how long star formation has been active in the region (duration of star formation, $t_{\rm SF}$). This assumption is also justified by studies of star-forming region complexes in local spirals \citep[e.g.][and references therein]{bastian05}. In the case of a region hosting only one (or more) WRs, we assume a fiducial age of 4~Myr. In general, we find that only 35 \% of the regions with $f_{\rm esc} >0$ have $t_{\rm SF}$ larger than the age of the youngest hosted clusters. This result is driven by the fact that most of the regions host single clusters (hence, $t_{\rm SF}$=$t_{\rm min}$ by definition).

\par Finally, we do not see strong dependencies between the $Q$ ratio and the total mass in YSCs (bottom plot in Fig.~\ref{fig:qratio_vs_age}, $r = 0.08$), suggesting that while star cluster mass seems to have a strong impact on the pressure terms at work within \hii regions, it does not drive the escape of ionising radiation. Similarly, we do not observe dependencies between the $Q$ ratio and $R_{\rm gc}$ (colour bar in the bottom plot of Fig.~\ref{fig:qratio_vs_age}). This suggests that none of the physical variations in the \hii regions which host star clusters is driving LyC escape from the regions. Taking into account that the observed $Q$ ratio $\ge 1$ would somewhat be affected by dust reprocessed light, especially at the youngest ages, these results point toward the scenario where star clusters are not the main responsible for LyC radiation leakage from \hii regions in M83. We discuss this further in Sect.~\ref{section:discussion}.

\section{Discussion}
\label{section:discussion}

\subsection{Pressure terms as function of YSC properties}
In the past decades, HST narrow-band imaging observations centred on H$\alpha$ emission, combined with broadband photometry of YSCs have been widely employed to study the evolution of \hii regions in nearby galaxies. Various studies find for example that after being initially embedded in their natal gas, clusters emerge after $<$ 4 -- 5~Myr, and according to the latest optically centred analyses \citep[][Hannon et al. 2022]{whitmore11, hollyhead15, hannon19, grasha19, messa21} this process can be as fast as 2 -- 3~Myr.
Whereas high-resolution imaging can inform us about the evolution of the \hii regions associated with star cluster properties, spectroscopy provides the key to actually sample \hii regions physical properties while they evolve.
Combining HST imaging of YSCs with MUSE spectroscopy of the ionised gas, we observed that indeed the properties of star clusters have a strong coupling with the pressure terms tracing stellar feedback exerted in \hii regions (Fig.~\ref{fig:P_vs_ysc}), and this despite the fact that in M83 YSCs only make up a fraction of the stars forming in the galaxy ($<$ 30\%). We suggest that  these trends are driven by the effect of clustering, i.e. the fact that feedback from stars tightly packed in clusters (having a size of a few parsecs) has a higher impact on the host \hii region, as observed in numerical simulations \citep[e.g.][]{kim17, gentry17, fielding18, bending20}.

\subsection{Relative strength of the pressure terms}
In Fig.~\ref{fig:P_vs_r}, we found that $P_{\rm ion}$ is predominant over $P_{\rm dir}$, with a median $P_{\rm ion}/P_{\rm dir} \simeq$ 13, but with significant variations in specific regions of the galaxy (like the central starburst at $R_e\leq 0.15$). These results are in agreement with what is reported in the literature for lower mass and metallicity galaxies (Large and Small Magellanic Cloud, NGC~300 at 2 Mpc), thus extending the range of galactic environment to massive, highly star-forming and metal-rich grand-design spiral galaxies. 
\citet{lopez11} studied the 30 Dor giant \hii region ($\sim$ 300 pc in size) in the Large Magellanic Cloud, finding that $P_{\rm ion}$ is predominant over $P_{\rm dir}$ at $d > 75$ pc from the central star cluster.
\citet{lopez14} performed a similar study in a sample of $\sim$ 30 \hii regions in the Small and Large Magellanic clouds, finding that $P_{\rm dir}$ is one or two order of magnitude smaller than $P_{\rm ion}$ in all regions. More recently, \citet{mcleod20, mcleod21} reported a median $P_{\rm ion}/P_{\rm dir} \sim $ 60 in a sample of $\sim$ 100 \hii regions in NGC~300.

\par In another recent study, \citet{barnes21} computed pressure terms for $\sim 5800$ regions in the PHANGS-MUSE sample, which consists of 19 spiral galaxies sampled from the local main-sequence of galaxies \citep{emsellem22} at an average distance of 15 Mpc. According to their selection criteria, M83 would nicely fit the sample. The authors obtain an upper and a lower limit for each pressure term, by considering the extreme cases of a perfectly smooth \hii region (with a pressure $P_{\rm min}$) and a clumpy \hii region, where all the clumps are located near the region's centre (with a pressure $P_{\rm max}$). Whereas in the upper limit case $P_{\rm dir}$ is observed to be $\sim P_{\rm ion}$, in the lower limit case $P_{\rm ion}$ is up to a factor four higher.

\par In order to better compare our results with the study of \citet{barnes21}, here we very briefly compare the adopted methodologies. For \citet{barnes21},  the \textit{lower limit} case the region radius corresponds to the measured effective radius ($R_e$), while the lower limit for $n_e$ is computed as a function of $Q_{\rm obs}$ and $R_e$.
In the \textit{upper limit} case, $n_e$ is estimated from the [\ion{S}{ii}] line ratio and the radius of the clumpy region is derived as function of $Q_{\rm obs}$ and $n_e$. In this work, on the other hand, we do not attempt to compute a lower and an upper limit. Our \hii region sizes are comparable to the radii estimated for the lower limit case in \citet{barnes21}, and agree with the size distribution of \hii regions observed in NGC~300 by \citet[][$\sim$ 10 pc resolution]{mcleod21}. We point out that, even if the sizes in Fig.~\ref{fig:hii_size_lum_distr} are overestimated by a factor three, $P_{\rm ion}$  would still be the dominant term.
Secondly, we compute pressure terms only for regions for which we have a density estimated from the [\ion{S}{ii}] line ratio (this approach removes about 30 regions from the 4684 extracted). Therefore, the luminosity, size and density of the regions are computed independently, and our estimates of $P_{\rm ion}$ and $P_{\rm dir}$ are independent from each other. Overall, our results are closer to those that \citet{barnes21} obtain by using the lower limit case (smooth \hii region assumption), reinforcing the conclusion that, for average star formation conditions representative of galaxies in the local universe, pre-SN feedback within \hii region is dominated by the thermal pressure exerted by photoionisation. 

\subsection{Pressure terms as function of galactocentric radius}
In Fig.~\ref{fig:P_vs_r}, we showed that -- outside of the inner region -- $P_{\rm dir}$ is approximately constant with radius, whereas $P_{\rm ion}$ decreases.
The pressure terms appear to be correlated both with changes in the intrinsic properties of the \hii regions (metallicity and thus radiation field strength, and extinction) and with changes in the galactic environmental conditions. While the trend in $P_{\rm dir}$ can be explained purely with changes in the region intrinsic properties, the decrease in $P_{\rm ion}$ with radius seems to be fully dominated by changes in the galactic environment. 
This is in agreement with what observed by \citet{barnes21}, where the pressure (both in their lower and upper limit) shows a systematic increase towards the galaxy centres, despite the higher metallicities.
On the other hand, \citet{mcleod21} find that in NGC~300 both $P_{\rm dir}$ and $P_{\rm ion}$ mildly increase with radius, and link these trends to the negative metallicity gradient and positive extinction gradient observed in the galaxy.

We suggest that this could be due either to the overall lower metallicity of NGC~300 with respect the metallicity measured in M83 and the majority of the PHANGS-MUSE galaxies, or to the fact that the environmental conditions are not significantly changing across the NGC~300 disk.

\par In order to study the dependency on the environment, adopting an approach similar to \citet{barnes21}, we compare the pressure of the \hii regions with the pressure of their surrounding environment, namely the midplane pressure measured at different distance from the centre of the galaxy.
We use the environmental pressure $P_\mathrm{DE}$ derived by \cite{sun2022} following the definition in \citet{leroy08} and \citet{sun20a}:
$$
P_\mathrm{DE} = \frac{\pi G}{2}\Sigma_\mathrm{gas}^2 + \Sigma_\mathrm{gas} \sqrt{2 G \rho_\star} \sigma_\mathrm{gas,\,z}~,
$$
where $\Sigma_\mathrm{gas}=\Sigma_\mathrm{mol}+\Sigma_\mathrm{atom}$ is the total gas surface density, $\rho_\star$ the stellar mass volume density near the galaxy mid-plane, and $\sigma_\mathrm{gas,\,z}$ the vertical velocity dispersion of the gas.
The values of $P_\mathrm{DE}$ reported by \cite{sun2022} were estimated within 500~pc wide radial bins across M83, using PHANGS-ALMA CO$\,(2{-}1)$ data \citep{leroy21_pipeline, leroy21}, THINGS HI data \citep{walter08}, and S$^4$G IRAC 3.6$\,\mu$m data \citep{sheth10}. There, $\Sigma_\mathrm{mol}$ was derived from CO line intensity with a radially varying CO-to-H$_2$ conversion factor \citep[following][]{sun20a}. A stellar mass surface density $\Sigma_\mathrm{\star}$ was derived from 3.6$\,\mu$m SB with a varying stellar mass-to-light ratio \citep[following][]{leroy21}, and then it was converted to $\rho_\mathrm{\star}$ with an estimated stellar disk scale height from the disk radial scale length $H_\star = R_\star / 7.3$ \citep{vanderkruit81, sun20a}.
A fixed gas velocity dispersion of $\sigma_\mathrm{gas,\,z}=10\;\mathrm{km\,s^{-1}}$ was adopted following \citet{leroy08}.
We remark that some of the assumptions that went into the calculation of $P_\mathrm{DE}$ might not hold well in the circumnuclear star-forming regions (innermost radial bin), where the stellar disk scale height could get larger, whereas the CO-to-H$_2$ conversion factor might be smaller than the adopted value.

\par
\begin{figure}
    \centering
    \includegraphics[width=\linewidth]{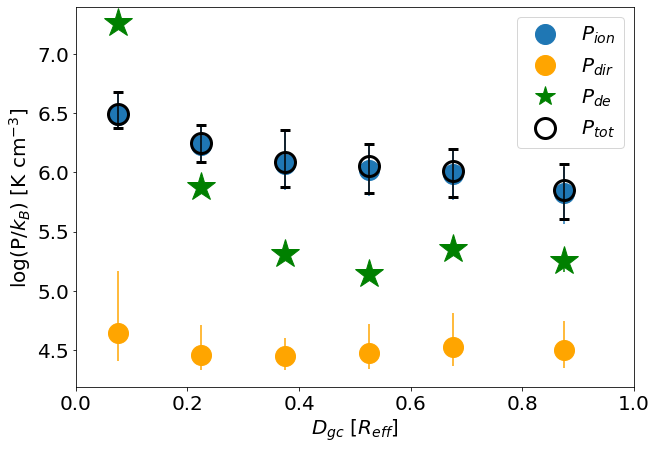}
    \caption{Comparison between the \hii region internal pressure terms ($P_{\rm ion}$ and $P_{\rm dir}$, in blue and orange, and their sum in black) and the environmental pressure ($P_\mathrm{DE}$, green stars). The pressure terms are binned in radial bins of width 0.15~$R_e$.}
    \label{fig:P_comparision}
\end{figure}

 In Fig.~\ref{fig:P_comparision} we compare $P_\mathrm{DE}$ with the internal pressure terms $P_{\rm ion}$ and $P_{\rm dir}$. We average all pressure terms in six radial bins of width 0.15~$R_e$. The resulting median $P_{\rm dir}$ and $P_{\rm ion}$ are shown as orange and blue filled circles, respectively, and their sum is shown as black open circles. The median $P_\mathrm{DE}$ are shown as green stars. We observe that $P_\mathrm{DE}$ increases dramatically towards the galactic center, where both the gas and stellar surface densities are higher. Its increase is strongly connected with the increase in the \hii region internal pressure terms. As a consequence, near the center the total internal pressure is lower than the environmental pressure, i.e., \hii regions are underpressured, whereas in the rest of the disk the regions are overpressured, and are therefore expanding. This is in agreement with what reported by \citet{barnes21} from the PHANGS data. The authors conclude that -- both in their lower and upper limit estimate --  increases in environmental pressure are driving the change in pressure terms. Namely, $n_e$ increases by up to an order of magnitude (as indicated in the bottom panel of Fig.~\ref{fig:hii_ebv_z_ne}), whereas the change in metallicity has a smaller impact on $T_e$.

\par In the literature, the physical conditions of the ISM in the central region of spiral galaxies such as the Milky Way or M83, have often been compared to the star formation conditions of main-sequence galaxies at the peak of the cosmic star formation history \citep[redshift $\sim$ 2,][]{kruijssen2013,ginsburg2019, callanan2021}. In light of these similarities, our results  suggest that pre-SN feedback drives the expansion of \hii regions, and therefore plays a significant role in dissolving the parent GMC in galactic environments where the star formation conditions are representative of galaxies in the local universe \citep[e.g. the galactic disk in M83, see also][]{chevance2022}. On the other hand, this might not be the case in galactic environments where the gas conditions probe extreme pressure and turbulence \citep[e.g.][]{leroy15,leroy2015a, callanan2021}, as is often seen  in high-redshift clumpy disk galaxies \citep{DZ2019, tacconi2020ARA&A}.
In the circumnuclear starburst region of M83, \hii regions appear to be underpressured and therefore stable against expansion. A similar behaviour has also been observed in \hii regions around massive star clusters in M82 \citep{smith06}, where $P_{\rm ion}/k_B$ has measured values of $\sim 10^7$ K cm$^{-3}$.

\par Unlike \cite{barnes21}, we have not attempted to estimate the pressure exerted by stellar winds, which do contribute to the total pressure exerted by pre-SN feedback in the early evolution stages of \hii regions. However, winds are unlikely to be a dominant factor in the expansion of the \hii regions \citep[e.g.][]{dale14}.
Overall, \citet{barnes21} find that $P_{\rm wind}$ is comparable to $P_{\rm dir}$ under the assumption of smooth \hii regions, or comparable to the $P_{\rm ion}$ term under the assumption of clumpy \hii regions. If we extrapolate these considerations for the \hii regions in the nuclear regions of M83 -- even in the extreme assumption of doubling the $P_{\rm ion}$ term to account for the missing $P_{\rm wind}$ -- we would get an average $P_{\rm tot}$ that would be (within uncertainties) still comparable to the midplane pressure. This suggests that pre-SN feedback might not be sufficient to drive the expansion and evolution of \hii regions in galaxies (or regions of galaxies) with more extreme gas conditions, both in the local universe as well as in high-redshift, gas-rich and highly turbulent disk galaxies.

\subsection{Ionisation budget}
In the ionisation budget analysis in Sect.~\ref{section:budget}, we found that ionising radiation escaped from only $\sim$ 13\% of the \hii regions. The large majority of them appears instead to be `missing' ionising photons. Dust can also mimic escape fraction, since it reprocesses LyC photons to emission at longer wavelengths. However, we do not see evidence that the \hii regions with $Q$ ratios higher than 1 have higher extinction in general. As a result, we conclude that the energy input by YSCs and WR stars alone is insufficient to explain the H$\alpha$ emission, both within and outside the \hii regions (in the DIG, see \paperI).

\par However, it is well known that very young star clusters are rarely isolated. As found in the analysis  of star-forming regions in the Local Group \citep[][and references therein]{gouliermis2018}, these clusters always sit in stellar overdensities; young, massive stars form around gravitationally bound clusters. Thus, to fully determine if there is escape of LyC photons from the \hii regions, we would also need to account for the contribution of the resolved stellar population surrounding the clusters. This is in apparent contrast with the results of \citet{weilbacher18}, who compared the \hii region emission and star cluster population in the Antennae galaxy merger. \citet{weilbacher18} found that $\sim$ 20\% of the regions are leaking LyC photons, and that leakage from the \hii regions was sufficient to explain the DIG observed in the system. However, most regions leaking LyC photons were found to populate the center of the merger, indicating that environmental conditions are playing a role.
\par Finally, our analysis does not show any direct relation between star cluster mass and escape fraction of LyC (bottom panel in Fig.~\ref{fig:qratio_vs_age}), although in the high mass range (m$_{tot} > 10^4$ \msun) between 30 and 50\% of the regions hosting YSC leak ionising continuum radiation. The low number statistics do not allow us to draw any conclusion. A way forward to test whether regions forming higher mass clusters favour escape of ionising radiation is to extend this type of analysis to other local spirals with elevated SFR to ensure an increase in the number of massive star clusters sampled. 

\par In the near future, the NIR and MIR wavelength that will be imaged by the James Webb Space Telescope (JWST) will allow us to trace emission reprocessed by dust (e.g. FEAST program, PI Adamo, ID 1793; the public treasury program ID 2107, PI Lee; among several others). We will be able to study the early phases of star formation and interaction with the parent GMC; account for the amount of LyC photons absorbed by dust, to improve cluster physical properties constraints in very young star clusters, as well as to account for the physical properties of very embedded \hii regions that are found to have high nebular extinction \citep[e.g.][]{messa21}. The large variety of galactic environments that will be probed by different programs will also enable us to study environmental dependencies and constraint the time scales and stellar feedback processes that are relevant for \hii region evolution and GMC dissolution. These time scales are fundamental because they are the link between stellar feedback originating at parsec scale and galactic scale properties of the gas that will regulate the future generation of star formation. 

\section{Summary and Conclusions}
\label{section:conclusions}
We present the analysis of $\sim$ 4700 \hii regions observed with MUSE across the disk of M83.
We identify the \hii regions based on their \ha\ emission. We then cross-match the initial catalogue with a sample of 148 PNe identified in the MUSE dataset (listed in Table~\ref{table:PN}), to  assess potential contaminant sources. We also match the \hii regions with the SNR catalogue of \citet{long2022}, finding that 149 regions are evolved regions hosting a SNR.

\par We compare the emission of the \hii regions with the stellar population they host. We spectroscopically identify in the MUSE data a total of 68 WR stars (listed in Table~\ref{table:WR_cc}).
We furthermore complement our data with a sample of $\sim$ 7500 YSCs observed with HST \citep{silva-villa14, adamo15}. In total, we had access to the star cluster and WR star population of $\sim$ 540 \hii regions.

\par We study the overall properties of the regions, finding that regions hosting YSCs are on average larger and more luminous (Fig.~\ref{fig:hii_size_lum_distr}), but that they otherwise have properties comparable to the rest of the sample. The average extinction peaks within the starburst region (at $R \sim 0.14~R_e$) and then decreases with radius (Fig.~\ref{fig:hii_ebv_z_ne}, top panel). A similar trend also exists in in oxygen abundance, and  we obtain a gradient in very good agreement with that determined by \citet[][Fig.~\ref{fig:hii_ebv_z_ne}, middle panel]{bresolin16}. Consequently, we find that the hardness of the radiation field increases with radius (Fig.~\ref{fig:hii_s3s2}). Also the electron density decreases with radius, and is $\simeq$ 2 dex higher in the central starburst region ($R \leq 0.15~R_e$) than in the rest of the disk (Fig.~\ref{fig:hii_ebv_z_ne}, bottom panel). Finally, the location of the \hii regions in a `BPT' emission line diagram confirms that they are compatible with pure photoionisation (Fig.~\ref{fig:hii_bpt}).

\par We investigate the impact of two feedback mechanisms on the regions: the pressure exerted by the ionised gas ($P_{\rm ion}$) and directly from the radiation ($P_{\rm dir}$). Overall, $P_{\rm ion}$ dominates over $P_{\rm dir}$ (Fig.~\ref{fig:P_vs_r}), in agreement with observations in the literature. Both internal pressure terms are enhanced in the nuclear region of the galaxy. The relative increase in $P_{\rm dir}$ is the strongest ($\simeq$ 2 dex). Outside the central region, $P_{\rm dir}$ stays approximately constant with radius: we interpret this as a combination of two factors: 1) a decrease in dust content (lowering the pressure) and 2) an increase in radiation field hardness (enhancing it), as shown in Fig.~\ref{fig:P_vs_ebv_Z}. On the other hand, $P_{\rm ion}$ decreases with radius. In this case, we explain the trend with a decrease in dust content (pointing to more evolved and hence less compact regions), and with the impact of the galactic environment. We investigate this further in Fig.~\ref{fig:P_comparision}, by comparing the internal pressure to the environmental pressure $P_\mathrm{DE}$ \citep{sun2022}. We find that regions in the galactic disk are over pressured and therefore expanding. On the other hand, for regions near the central starburst $P_\mathrm{DE}$ is almost one order of magnitude higher than the internal pressure, pointing to the fact that in extreme gas conditions pre-SN feedback is not sufficient to drive \hii region expansion.

\par We constrain the age, mass and ionising photon flux of the YSCs based on their photometry, using a Bayesian fitting analysis and stochastically populated cluster models.
We observe that YSCs populating the \hii regions are on average
younger (median age of 4.7 vs 8.2 Myr), more massive (median mass of $1.7$ vs $1.4 \times 10^3$~\msun), and emit an order of magnitude more ionising photons (median $\log Q(\mbox{H}^0)$ of 48.5 vs 47.4 s$^{-1}$) than field clusters. We then investigate how the pressure terms are impacted by the stellar population powering the regions. We find that regions hosting younger clusters have higher internal pressure, peaking at $t_{\rm min, YSC} \simeq 1$~Myr (Fig.~\ref{fig:P_vs_ysc}).

Whereas $P_{\rm ion}$ flattens after 2~Myr, the decreasing trend in $P_{\rm dir}$ continues up to 10~Myr, suggesting that the coupling between the pressure terms and the \hii regions grows steadily weaker with age.
Regions hosting a higher mass in young clusters ($t \leq $ 10~Myr) have a higher internal pressure, with a more pronounced trend for $P_{\rm dir}$. These trends have never been observed before, and seem to suggest that young, clustered star formation has a stronger impact on the \hii regions than distributed star formation, in agreement with the results of numerical simulations. This seems to hold despite the fact that the fraction of stars formed in compact, gravitationally-bound clusters in M83 is low \citep[$<$ 30\%,][]{adamo15}.

\par Finally, we have computed a photoionisation budget for the regions for which we have access to the stellar population (limiting our analysis to the YSCs and WR stars). We compare the modeled flux to the observed H$\alpha$ emission to compute an escape fraction. Overall, we detect escaping radiation in only  $\sim$ 13\% of the regions. The majority of these regions hosts cluster younger than 2 Myr. The $f_{\rm esc}$ does not seem to positively correlate either with the duration of star formation in the regions, or with the total mass in young clusters (Fig.~\ref{fig:qratio_vs_age}). We conclude that the energy input of YSCs and WR stars by itself is not sufficient to explain the observed \ha\ emission, both within the regions and in the DIG. Further analysis, including the assessment of the resolved stellar population surrounding each cluster, is needed to complete the physical picture.

\begin{acknowledgements}
We thank the anonymous referee for suggestions and comments that have improved the manuscript. A.~A. acknowledges support of the Swedish Research Council, Vetenskapsr\aa{}det (2016-05199), and the Swedish National Space Agency (SNSA, Dnr140).
The work of J.~S. is partially supported by the Natural Sciences and Engineering Research Council of Canada (NSERC) through the Canadian Institute for Theoretical Astrophysics (CITA) National Fellowship, and by the National Science Foundation (NSF) under Grants No. 1615105, 1615109, and 1653300.
This work is based on observations collected at the European Southern Observatory under ESO programmes
096.B-0057(A), 0101.B-0727(A),
097.B-0899(B), 
097.B-0640(A). 
This research made use of Astropy\footnote{http://www.astropy.org}, a community-developed core Python package for Astronomy \citep{astropy:2013, astropy:2018}. 
\end{acknowledgements}

\bibpunct{(}{)}{;}{a}{}{,} 
\bibliographystyle{aa}
\bibliography{paperIV}

\begin{appendix}
\onecolumn

\section{\hii region selection}
\label{section:appendix_HII}
\begin{figure*}[!htb]
    \centering
    \includegraphics[width=0.32\linewidth]{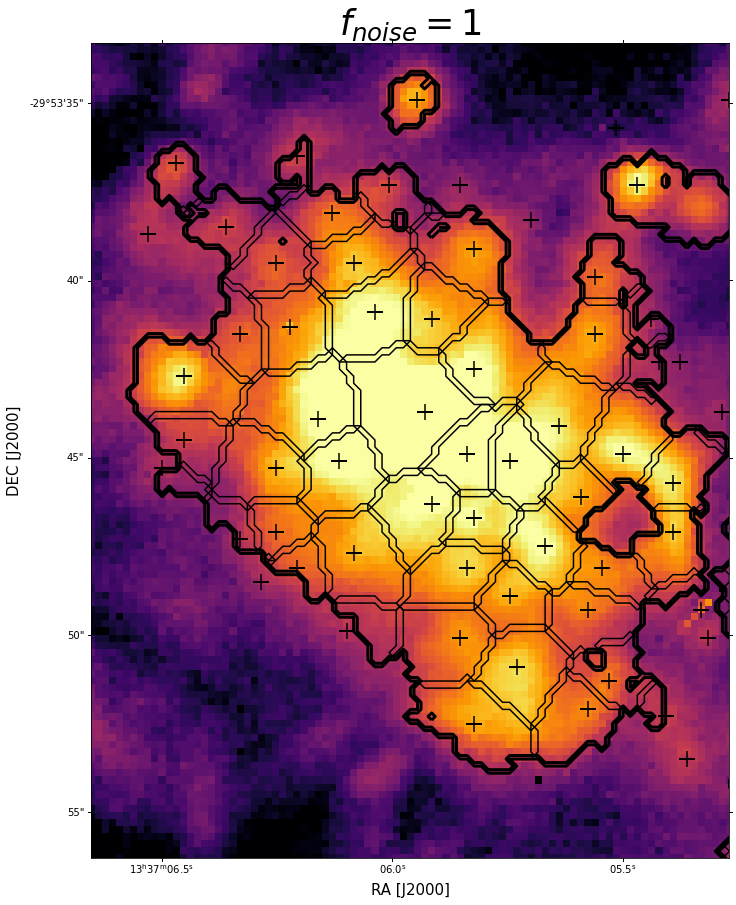}
    \includegraphics[width=0.31\linewidth]{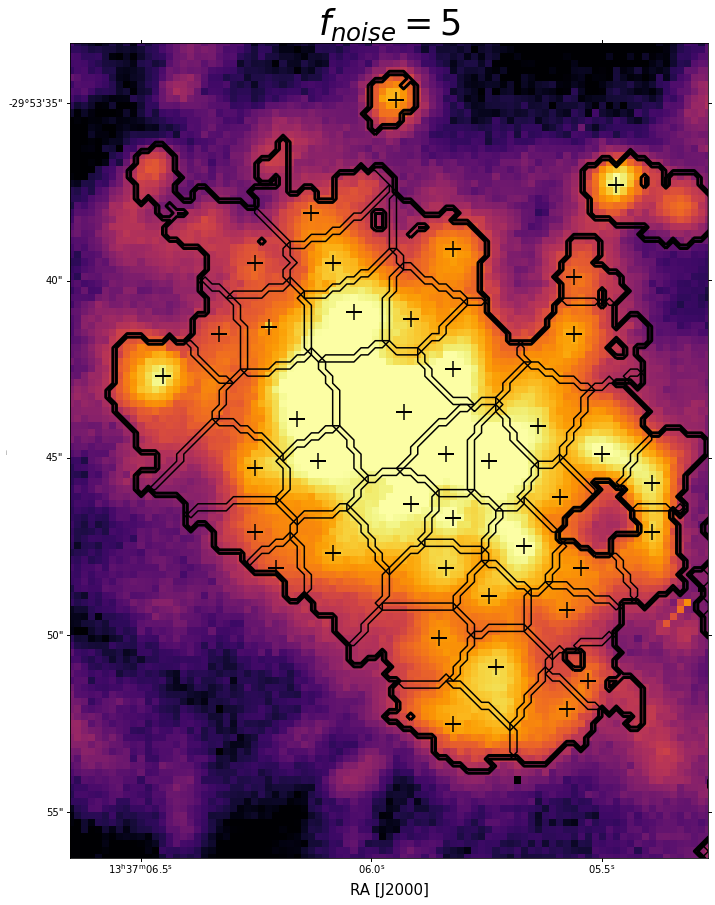}
    \includegraphics[width=0.31\linewidth]{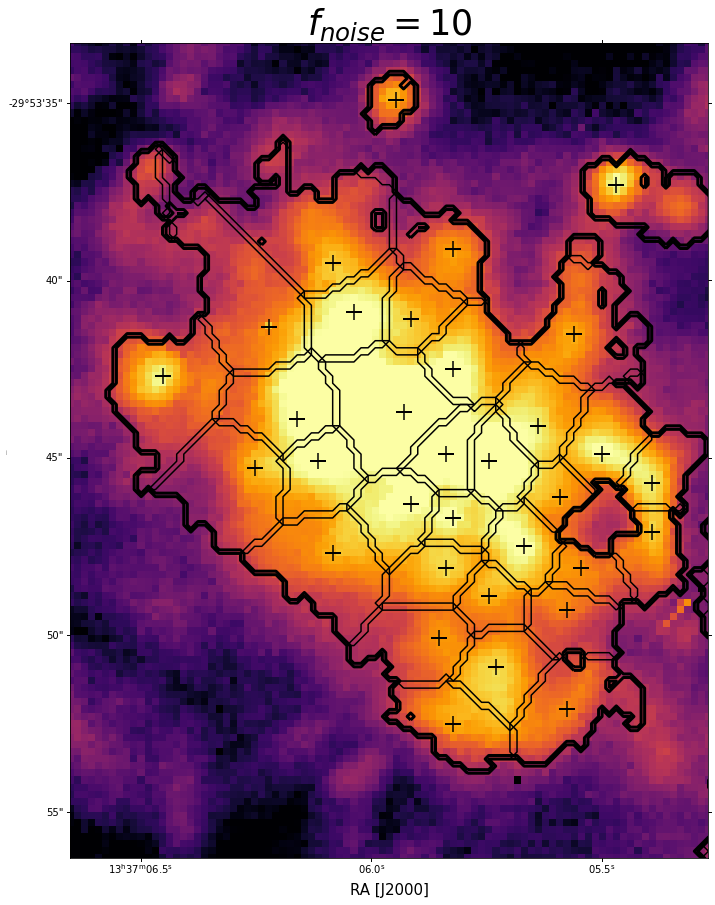}
    \caption{Emission peaks (black `{\Large{+}}') and boundaries (thin black lines) of regions identified in a large \hii complex with three different values of the $f_{\rm noise}$ parameter (Eq.~\ref{eq:hii_threshold}). The thick black contours indicate the SB threshold adopted for the \hii regions (see Sect.~\ref{section:hii_identification}).}
    \label{fig:HII_contours_comparison}
\end{figure*}

In this appendix we expand on the \hii region selection procedure described in Sect.~\ref{section:hii_identification}. In particular, we show the impact of the $f_{\rm noise}$ parameter (see Eq.~\ref{eq:hii_threshold}) on the boundaries, luminosity and size of the recovered regions.
In Fig.~\ref{fig:HII_contours_comparison} we show the emission peaks (black `{\Large{+}}') and region boundaries (thin black lines) identified with three different values of $f_{\rm noise}$.
By visual inspection, we find that a value of $f_{\rm noise} = 1$ (left panel) delivers an optimal division of the regions, as it detects all relevant peaks of emission within the adopted SB limit (thick black contour in Fig.~\ref{fig:HII_contours_comparison}).
We find that values of $f_{\rm noise} < 1$ do not change the number and distribution of peaks within the SB contours, but only lead to the detection of additional fainter peaks.
A progressive increase of $f_{\rm noise}$ (central and right panel in Fig.~\ref{fig:HII_contours_comparison}), on the other hand, results in the missed detection of relevant emission peaks and therefore in unrealistic region boundaries.

\par
\begin{figure*}[!htb]
    \centering
    \includegraphics[width=0.45\linewidth]{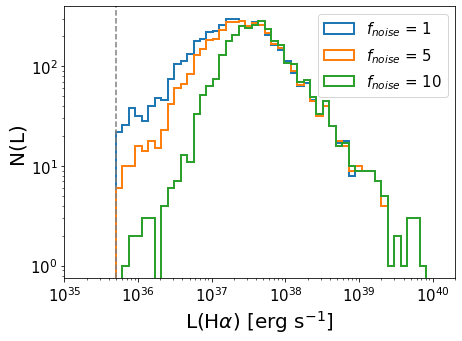}
    \includegraphics[width=0.45\linewidth]{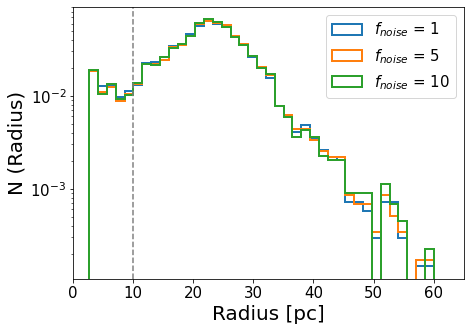}
    \caption{Luminosity function (\textit{left panel}) and size distribution (\textit{right panel}) for the samples of \hii regions obtained with three different values of the $f_{\rm noise}$ parameter (Eq.~\ref{eq:hii_threshold}). The grey dashed lines indicate, respectively, the adopted adopted \hii SB threshold and the spatial resolution limit of the MUSE data.}
    \label{fig:HII_Lum_size_distr_comparison}
\end{figure*}

 Fig.~\ref{fig:HII_Lum_size_distr_comparison} illustrates the luminosity function (left panel) and size distribution (right panel) of the three resulting samples of \hii regions (across the full disk). In the luminosity function, we observe that the distributions match for $L(H\alpha) > 5 \times 10^{37}$, but that with increasing $f_{\rm noise}$ the histogram cuts off at a progressively higher threshold. The size distributions, on the other hand, match well at all radii.

\section{PNe sample}
\label{section:appendix_pn}

In this appendix we provide some complementary information related to the PNe detection and confirmation. As discussed in Sect.~\ref{section:pn}, we use two different methods to detect PNe candidates. In the first method, we use  the [O~III]/H$\alpha$ map extracted from the MUSE mosaic. In the second method, we complement the purely spectroscopic based selection by visually inspecting HST  images. The second method enables us to detect PNe candidates that would otherwise be missed in our initial selection. We use the position of these targets to extract their spectra and verify whether their line ratios would confirm their nature or not. We find an additional 28 PNe that are spectroscopically confirmed and are therefore included in the confirmed sample. The remaining targets are retained as candidate  PNe. Unfortunately the  comparatively low spatial resolution of the MUSE data does not allow us to fully confirm the nature of the latter systems. 

\begin{figure*}[!htb]
    \centering
    \includegraphics[width=19cm]{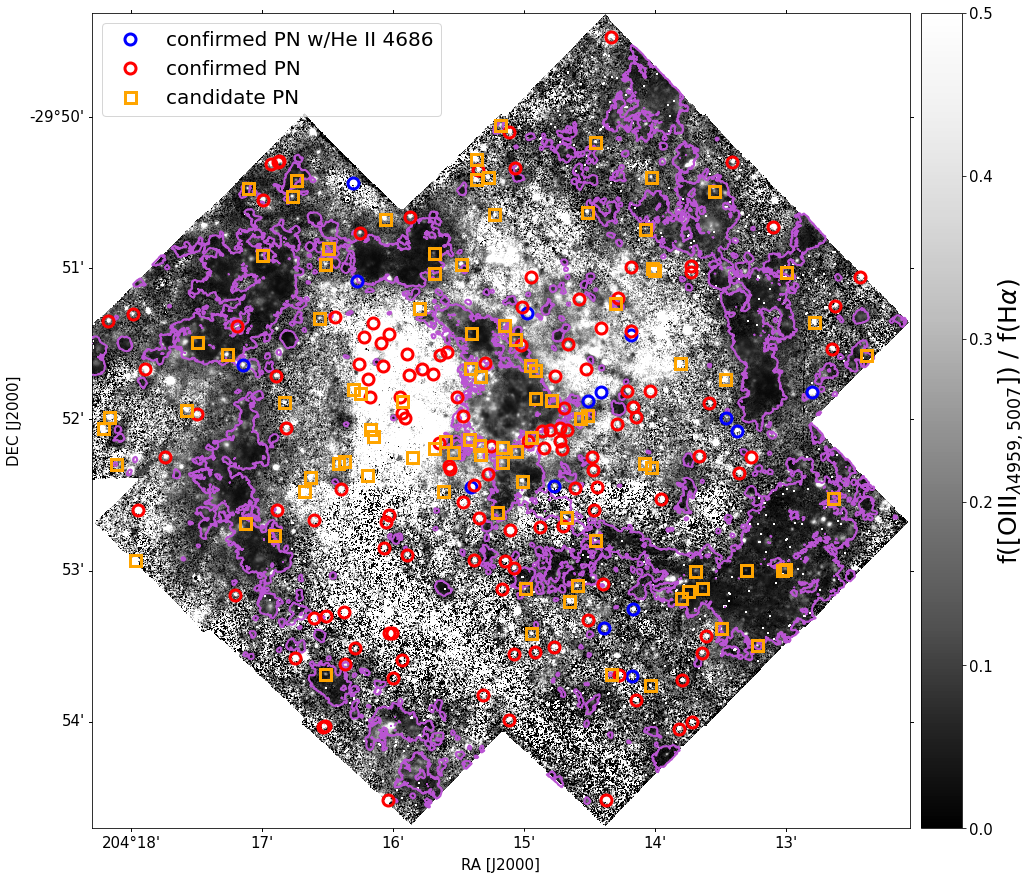}
    \caption{Location of the identified PNe on a map of [\ion{O}{iii}]/H$\alpha$. The coordinates of all objects are given in Table~\ref{table:PN}. Blue and red circles indicate spectroscopically confirmed candidates (Class 1 and 2 in Table~\ref{table:PN}) with and without \ion{He}{II}~$\lambda$4686 detection, respectively. The orange squares indicate candidates visually identified in the HST+MUSE dataset (Class 3 in Table~\ref{table:PN}). In purple we indicate the outer boundaries of the \hii regions.
    }
    \label{fig:PN}
\end{figure*}

\begin{figure*}[!htb]
    \centering
    \includegraphics[width=\linewidth]{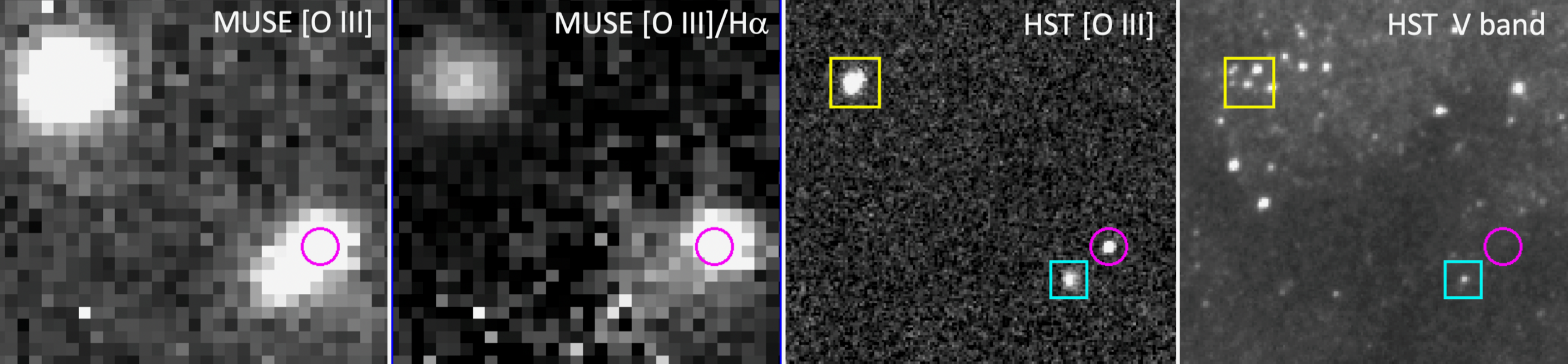}
     \caption{An example of the visual inspection technique applied to a small 6\arcsec\ region of M83. The small magenta circle (0.6\arcsec\ diameter) shows a confirmed PN candidate, which appears as a point-like source of [\ion{O}{iii}] emission in the HST data.  The yellow square shows an object that looks very similar in the MUSE data but is clearly extended in the HST data, thus disqualifying it from further consideration.  Likewise, the cyan square shows an object with [\ion{O}{iii}] emission, which however aligns with a stellar source in V band.  This demonstrates the power of using MUSE and WFC3 together to identify candidate PN at the distance of M83.}
     \label{fig:pn_visual_inspection}
\end{figure*}

We show an example of the search technique applied to a small 6\arcsec\ region of M83 in Fig.~\ref{fig:pn_visual_inspection}. The positions of all spectroscopically confirmed and candidate PNe are shown in Fig.~\ref{fig:PN}. Their coordinates and detection class are listed in Table~\ref{table:PN}.

\newpage

\topcaption{ Coordinates of the PNe identified in the MUSE+HST dataset. In the third column we report a classification score, indicating whether the PNe is spectroscopically confirmed (class 1, additional \ion{He}{ii}$~\lambda$4686 detection, and 2), or whether it remains a candidate PNe (class 3, visually identified). In the last column we cross reference our sample with PN catalogued by \citet{herrmann09}.}
\tablefirsthead{ \toprule \toprule
  Obj. ID &   Ra (J2000) &   Dec (J2000) &  Class. &  Herrmann ID \\ \midrule}
\tablehead{ 
\toprule \toprule
  Obj. ID &   Ra (J2000) &   Dec (J2000) &  Class. &  Herrmann ID \\ \midrule }
\begin{supertabular}{lllll}
PN1 & 13:36:49.569 & -29:51:34.311 & 3 &\\     
PN2 & 13:36:49.728 & -29:51:03.469 & 2 &\\     
PN3 & 13:36:50.506 & -29:51:14.808 & 2 &\\     
PN4 & 13:36:50.558 & -29:52:31.065 & 3 &\\     
PN5 & 13:36:50.601 & -29:51:31.951 & 2 &\\     
PN6 & 13:36:51.154 & -29:51:21.157 & 3 &\\     
PN7 & 13:36:51.216 & -29:51:49.212 & 1 &\\    
PN8 & 13:36:52.003 & -29:51:01.525 & 3 &\\     
PN9 & 13:36:52.039 & -29:52:59.310 & 3 &\\     
PN10 & 13:36:52.128 & -29:52:59.653 & 3 &\\     
PN11 & 13:36:52.383 & -29:50:43.551 & 2 &\\     
PN12 & 13:36:52.880 & -29:53:29.726 & 3 &\\     
PN13 & 13:36:53.060 & -29:52:14.952 & 2 &\\     
PN14 & 13:36:53.221 & -29:52:59.836 & 3 &\\     
PN15 & 13:36:53.445 & -29:52:21.200 & 2 &\\     
PN16 & 13:36:53.502 & -29:52:04.581 & 1 &\\    
PN17 & 13:36:53.637 & -29:50:17.641 & 2 &\\     
PN18 & 13:36:53.818 & -29:51:59.469 & 1 &\\    
PN19 & 13:36:53.870 & -29:51:43.980 & 3 &\\     
PN20 & 13:36:53.986 & -29:53:23.051 & 3 &\\     
PN21 & 13:36:54.194 & -29:50:29.420 & 3 &\\     
PN22 & 13:36:54.334 & -29:51:53.593 & 2 &\\     
PN23 & 13:36:54.431 & -29:53:26.186 & 2 &\\     
PN24 & 13:36:54.561 & -29:53:07.038 & 3 &\\     
PN25 & 13:36:54.573 & -29:53:32.735 & 2 &\\     
PN26 & 13:36:54.647 & -29:52:14.630 & 2 &\\     
PN27 & 13:36:54.785 & -29:53:00.410 & 3 &\\     
PN28 & 13:36:54.880 & -29:54:00.104 & 2 &\\     
PN29 & 13:36:54.886 & -29:50:58.973 & 2 &\\     
PN30 & 13:36:54.903 & -29:51:01.423 & 2 &\\     
PN31 & 13:36:54.980 & -29:53:08.254 & 3 &\\     
PN32 & 13:36:55.177 & -29:53:43.636 & 2 &\\     
PN33 & 13:36:55.187 & -29:53:11.076 & 3 &\\     
PN34 & 13:36:55.246 & -29:51:37.660 & 3 &\\     
PN35 & 13:36:55.275 & -29:54:02.799 & 2 &\\     
PN36 & 13:36:55.800 & -29:52:31.703 & 2 &\\     
PN37 & 13:36:56.015 & -29:51:00.618 & 3 &\\     
PN38 & 13:36:56.075 & -29:51:00.043 & 3 &\\     
PN39 & 13:36:56.113 & -29:52:19.090 & 3 &\\     
PN40 & 13:36:56.127 & -29:50:23.710 & 3 &\\     
PN41 & 13:36:56.140 & -29:51:48.646 & 2 &\\     
PN42 & 13:36:56.145 & -29:53:45.637 & 3 &\\     
PN43 & 13:36:56.301 & -29:50:44.352 & 3 &\\     
PN44 & 13:36:56.322 & -29:52:17.275 & 3 &\\     
PN45 & 13:36:56.571 & -29:53:51.506 & 2 &\\     
PN46 & 13:36:56.572 & -29:51:58.889 & 2 &\\     
PN47 & 13:36:56.665 & -29:53:15.496 & 1 &\\    
PN48 & 13:36:56.673 & -29:51:54.963 & 2 &\\     
PN49 & 13:36:56.708 & -29:53:41.756 & 1 &\\    
PN50 & 13:36:56.721 & -29:50:59.351 & 2 &\\     
PN51 & 13:36:56.723 & -29:51:24.724 & 2 &\\     
PN52 & 13:36:56.733 & -29:51:26.391 & 1 &\\    
PN53 & 13:36:56.846 & -29:51:48.816 & 2 &\\     
PN54 & 13:36:57.081 & -29:53:41.395 & 2 &\\     
PN55 & 13:36:57.134 & -29:51:11.706 & 2 &\\     
PN56 & 13:36:57.154 & -29:52:02.016 & 2 &\\     
PN57 & 13:36:57.206 & -29:51:13.619 & 3 &\\     
PN58 & 13:36:57.330 & -29:53:41.006 & 3 &\\     
PN59 & 13:36:57.350 & -29:49:28.122 & 2 &\\     
PN60 & 13:36:57.495 & -29:54:31.075 & 2 &\\     
PN61 & 13:36:57.553 & -29:53:22.925 & 1 &\\    
PN62 & 13:36:57.575 & -29:53:05.537 & 2 &\\     
PN63 & 13:36:57.631 & -29:51:49.210 & 1 &\\    
PN64 & 13:36:57.660 & -29:51:23.798 & 2 &\\     
PN65 & 13:36:57.759 & -29:52:26.905 & 2 &\\     
PN66 & 13:36:57.820 & -29:50:09.776 & 3 &\\     
PN67 & 13:36:57.821 & -29:52:48.103 & 3 &\\     
PN68 & 13:36:57.857 & -29:52:36.020 & 2 &\\     
PN69 & 13:36:57.899 & -29:52:19.993 & 2 &\\     
PN70 & 13:36:57.931 & -29:52:14.995 & 2 &\\     
PN71 & 13:36:58.029 & -29:53:19.822 & 2 &\\     
PN72 & 13:36:58.057 & -29:51:52.676 & 1 &\\    
PN73 & 13:36:58.066 & -29:51:58.095 & 3 &\\     
PN74 & 13:36:58.073 & -29:50:37.521 & 3 &\\     
PN75 & 13:36:58.101 & -29:51:40.163 & 2 &\\     
PN76 & 13:36:58.299 & -29:51:59.662 & 3 &\\     
PN77 & 13:36:58.305 & -29:51:12.296 & 2 &\\     
PN78 & 13:36:58.384 & -29:53:05.939 & 3 &\\     
PN79 & 13:36:58.445 & -29:52:27.080 & 2 &\\     
PN80 & 13:36:58.622 & -29:53:11.956 & 3 &\\     
PN81 & 13:36:58.656 & -29:51:30.197 & 2 &\\     
PN82 & 13:36:58.676 & -29:52:04.308 & 2 &\\     
PN83 & 13:36:58.727 & -29:52:38.858 & 3 &\\     
PN84 & 13:36:58.788 & -29:51:55.667 & 2 &\\     
PN85 & 13:36:58.795 & -29:52:42.286 & 2 &\\     
PN86 & 13:36:58.840 & -29:52:11.790 & 2 &\\     
PN87 & 13:36:58.906 & -29:52:03.406 & 2 &\\     
PN88 & 13:36:58.907 & -29:52:08.480 & 2 &\\     
PN89 & 13:36:59.065 & -29:51:42.670 & 2 &\\     
PN90 & 13:36:59.078 & -29:53:30.366 & 2 &\\     
PN91 & 13:36:59.079 & -29:52:26.606 & 1 &\\    
PN92 & 13:36:59.165 & -29:51:52.484 & 3 &\\     
PN93 & 13:36:59.193 & -29:52:04.201 & 2 &\\     
PN94 & 13:36:59.391 & -29:52:11.335 & 2 &\\     
PN95 & 13:36:59.445 & -29:52:04.737 & 2 &\\     
PN96 & 13:36:59.508 & -29:52:42.680 & 2 &\\     
PN97 & 13:36:59.654 & -29:51:51.445 & 3 &\\     
PN98 & 13:36:59.659 & -29:51:40.538 & 3 &\\     
PN99 & 13:36:59.659 & -29:53:32.512 & 2 &\\     
PN100 & 13:36:59.773 & -29:53:24.805 & 3 &\\     
PN101 & 13:36:59.783 & -29:51:03.449 & 2 &\\     
PN102 & 13:36:59.793 & -29:52:07.125 & 3 &\\     
PN103 & 13:36:59.818 & -29:51:38.578 & 3 &\\     
PN104 & 13:36:59.877 & -29:52:08.422 & 2 &\\     
PN105 & 13:36:59.914 & -29:51:17.693 & 1 &\\    
PN106 & 13:36:59.972 & -29:53:07.079 & 3 &\\     
PN107 & 13:37:00.051 & -29:51:15.578 & 2 &\\     
PN108 & 13:37:00.060 & -29:52:24.347 & 3 &\\     
PN109 & 13:37:00.087 & -29:51:30.330 & 2 &\\     
PN110 & 13:37:00.250 & -29:52:12.396 & 3 &\\     
PN111 & 13:37:00.268 & -29:50:20.188 & 2 &\\     
PN112 & 13:37:00.283 & -29:51:28.250 & 3 &\\     
PN113 & 13:37:00.290 & -29:53:33.184 & 2 &\\     
PN114 & 13:37:00.314 & -29:52:59.147 & 2 & M83-111 \\
PN115 & 13:37:00.423 & -29:52:43.979 & 2 &\\     
PN116 & 13:37:00.447 & -29:50:05.868 & 2 &\\     
PN117 & 13:37:00.465 & -29:53:59.255 & 2 &\\     
PN118 & 13:37:00.563 & -29:52:56.106 & 2 &  M83-80 \\
PN119 & 13:37:00.591 & -29:51:22.418 & 3 &\\     
PN120 & 13:37:00.666 & -29:53:07.495 & 2 &\\     
PN121 & 13:37:00.668 & -29:52:10.879 & 3 &\\     
PN122 & 13:37:00.670 & -29:52:16.925 & 3 &\\     
PN123 & 13:37:00.726 & -29:50:03.099 & 3 &\\     
PN124 & 13:37:00.825 & -29:52:36.844 & 3 &\\     
PN125 & 13:37:00.898 & -29:50:38.614 & 3 &\\     
PN126 & 13:37:01.001 & -29:52:10.454 & 2 &\\     
PN127 & 13:37:01.090 & -29:50:23.919 & 3 &\\     
PN128 & 13:37:01.092 & -29:52:21.879 & 2 &\\     
PN129 & 13:37:01.199 & -29:51:37.491 & 2 &\\     
PN130 & 13:37:01.233 & -29:53:49.591 & 2 &\\     
PN131 & 13:37:01.335 & -29:51:42.897 & 3 &\\     
PN132 & 13:37:01.344 & -29:52:13.674 & 3 &\\     
PN133 & 13:37:01.360 & -29:52:39.358 & 2 &\\     
PN134 & 13:37:01.368 & -29:52:10.125 & 3 &\\     
PN135 & 13:37:01.396 & -29:50:20.993 & 2 &\\     
PN136 & 13:37:01.472 & -29:50:16.785 & 3 &\\     
PN137 & 13:37:01.472 & -29:50:24.649 & 3 &\\     
PN138 & 13:37:01.515 & -29:52:25.895 & 2 &\\     
PN139 & 13:37:01.525 & -29:52:56.028 & 2 &\\     
PN140 & 13:37:01.607 & -29:51:25.674 & 3 &\\     
PN141 & 13:37:01.616 & -29:52:27.020 & 1 &\\    
PN142 & 13:37:01.640 & -29:51:39.619 & 3 &\\     
PN143 & 13:37:01.678 & -29:52:07.897 & 3 &\\     
PN144 & 13:37:01.855 & -29:52:32.927 & 2 &\\     
PN145 & 13:37:01.870 & -29:51:58.850 & 2 &\\     
PN146 & 13:37:01.920 & -29:50:58.318 & 3 &\\     
PN147 & 13:37:02.043 & -29:51:50.976 & 2 &\\     
PN148 & 13:37:02.162 & -29:52:13.163 & 3 &\\     
PN149 & 13:37:02.263 & -29:52:19.705 & 2 &\\     
PN150 & 13:37:02.271 & -29:52:18.368 & 2 &\\     
PN151 & 13:37:02.341 & -29:51:33.101 & 2 &\\     
PN152 & 13:37:02.405 & -29:52:08.691 & 3 &\\     
PN153 & 13:37:02.457 & -29:52:28.447 & 3 &\\     
PN154 & 13:37:02.550 & -29:51:34.394 & 2 &\\     
PN155 & 13:37:02.583 & -29:52:09.310 & 2 &\\     
PN156 & 13:37:02.730 & -29:52:11.229 & 3 &\\     
PN157 & 13:37:02.748 & -29:51:02.071 & 3 &\\     
PN158 & 13:37:02.757 & -29:50:53.917 & 3 &\\     
PN159 & 13:37:02.787 & -29:51:42.189 & 2 &\\     
PN160 & 13:37:03.110 & -29:51:40.158 & 2 &\\     
PN161 & 13:37:03.208 & -29:51:15.888 & 3 &\\     
PN162 & 13:37:03.407 & -29:52:15.135 & 3 &\\     
PN163 & 13:37:03.481 & -29:50:39.560 & 2 &\\     
PN164 & 13:37:03.504 & -29:51:42.403 & 2 &\\     
PN165 & 13:37:03.575 & -29:51:33.859 & 2 &\\     
PN166 & 13:37:03.576 & -29:52:53.913 & 2 &\\     
PN167 & 13:37:03.627 & -29:51:59.488 & 2 &\\     
PN168 & 13:37:03.710 & -29:51:52.761 & 3 &\\     
PN169 & 13:37:03.717 & -29:51:57.439 & 2 &\\     
PN170 & 13:37:03.734 & -29:53:35.445 & 2 &\\     
PN171 & 13:37:03.768 & -29:51:50.996 & 2 &\\     
PN172 & 13:37:03.995 & -29:53:42.676 & 2 &\\     
PN173 & 13:37:04.026 & -29:53:24.868 & 2 &\\     
PN174 & 13:37:04.112 & -29:51:25.924 & 2 &\\     
PN175 & 13:37:04.124 & -29:53:24.984 & 2 &\\     
PN176 & 13:37:04.128 & -29:52:38.057 & 2 &\\     
PN177 & 13:37:04.163 & -29:54:31.086 & 2 & M83-7 \\
PN178 & 13:37:04.194 & -29:52:40.639 & 2 &\\     
PN179 & 13:37:04.252 & -29:50:40.317 & 3 &\\     
PN180 & 13:37:04.264 & -29:52:51.222 & 2 &\\     
PN181 & 13:37:04.286 & -29:51:38.819 & 2 &\\     
PN182 & 13:37:04.364 & -29:51:29.773 & 2 &\\     
PN183 & 13:37:04.600 & -29:52:06.666 & 3 &\\     
PN184 & 13:37:04.616 & -29:51:21.797 & 2 &\\     
PN185 & 13:37:04.690 & -29:51:51.262 & 2 & M83-11 \\
PN186 & 13:37:04.694 & -29:52:03.891 & 3 &\\     
PN187 & 13:37:04.770 & -29:51:43.806 & 2 &\\     
PN188 & 13:37:04.791 & -29:52:21.967 & 3 &\\     
PN189 & 13:37:04.867 & -29:51:27.367 & 2 &\\     
PN190 & 13:37:04.999 & -29:51:49.587 & 3 &\\     
PN191 & 13:37:05.006 & -29:50:45.824 & 2 &\\     
PN192 & 13:37:05.034 & -29:51:38.071 & 2 &\\     
PN193 & 13:37:05.088 & -29:51:04.960 & 1 &\\    
PN194 & 13:37:05.153 & -29:53:30.926 & 2 &\\     
PN195 & 13:37:05.206 & -29:50:25.993 & 1 & M83-89 \\
PN196 & 13:37:05.211 & -29:51:48.046 & 3 &\\     
PN197 & 13:37:05.449 & -29:53:37.008 & 2 &\\     
PN198 & 13:37:05.481 & -29:53:16.478 & 2 &\\     
PN199 & 13:37:05.491 & -29:52:16.539 & 3 &\\     
PN200 & 13:37:05.589 & -29:52:27.780 & 2 &\\     
PN201 & 13:37:05.684 & -29:52:17.482 & 3 &\\     
PN202 & 13:37:05.763 & -29:51:19.372 & 2 &\\     
PN203 & 13:37:05.990 & -29:50:51.946 & 3 &\\     
PN204 & 13:37:06.047 & -29:53:18.071 & 2 &\\     
PN205 & 13:37:06.057 & -29:53:41.073 & 3 &\\     
PN206 & 13:37:06.058 & -29:50:58.392 & 3 &\\     
PN207 & 13:37:06.065 & -29:54:01.788 & 2 &\\     
PN208 & 13:37:06.136 & -29:54:02.050 & 2 &\\     
PN209 & 13:37:06.246 & -29:51:19.651 & 3 &\\     
PN210 & 13:37:06.408 & -29:52:39.886 & 2 &\\     
PN211 & 13:37:06.420 & -29:53:18.691 & 2 &\\     
PN212 & 13:37:06.536 & -29:52:22.984 & 3 &\\     
PN213 & 13:37:06.702 & -29:52:28.604 & 3 &\\     
PN214 & 13:37:06.952 & -29:50:24.900 & 3 &\\     
PN215 & 13:37:06.984 & -29:53:34.932 & 2 &\\     
PN216 & 13:37:07.068 & -29:50:31.323 & 3 &\\     
PN217 & 13:37:07.260 & -29:52:03.360 & 2 &\\     
PN218 & 13:37:07.318 & -29:51:53.115 & 3 &\\     
PN219 & 13:37:07.459 & -29:50:17.248 & 2 &\\     
PN220 & 13:37:07.517 & -29:50:17.609 & 2 &\\     
PN221 & 13:37:07.526 & -29:52:36.061 & 2 &\\     
PN222 & 13:37:07.562 & -29:51:42.778 & 2 &\\     
PN223 & 13:37:07.632 & -29:52:46.065 & 3 &\\     
PN224 & 13:37:07.711 & -29:50:18.680 & 2 &\\     
PN225 & 13:37:07.969 & -29:50:32.859 & 2 &\\     
PN226 & 13:37:07.988 & -29:50:54.673 & 3 &\\     
PN227 & 13:37:08.430 & -29:50:28.278 & 3 &\\     
PN228 & 13:37:08.522 & -29:52:41.036 & 3 &\\     
PN229 & 13:37:08.585 & -29:51:38.487 & 1 &\\    
PN230 & 13:37:08.762 & -29:51:22.728 & 2 &\\     
PN231 & 13:37:08.811 & -29:53:09.666 & 2 &\\     
PN232 & 13:37:09.070 & -29:51:34.230 & 3 &\\     
PN233 & 13:37:09.989 & -29:51:58.002 & 2 &\\     
PN234 & 13:37:09.993 & -29:51:29.106 & 3 &\\     
PN235 & 13:37:10.319 & -29:51:56.265 & 3 & M83-100 \\
PN236 & 13:37:10.946 & -29:52:14.938 & 2 &\\     
PN237 & 13:37:11.560 & -29:51:39.862 & 2 &\\     
PN238 & 13:37:11.776 & -29:52:36.101 & 2 &\\     
PN239 & 13:37:11.859 & -29:52:55.856 & 3 & M83-195 \\
PN240 & 13:37:11.944 & -29:51:18.286 & 2 &\\     
PN241 & 13:37:12.464 & -29:52:17.532 & 3 &\\     
PN242 & 13:37:12.671 & -29:51:59.060 & 3 &\\     
PN243 & 13:37:12.681 & -29:51:20.933 & 2 &\\     
PN244 & 13:37:12.859 & -29:52:03.302 & 3 &\\     
\end{supertabular}
\label{table:PN}

\newpage
\section{WR sample}
\label{section:appendix_wr}
\begin{figure*}[!htb]
    \centering
    \includegraphics[width=19cm]{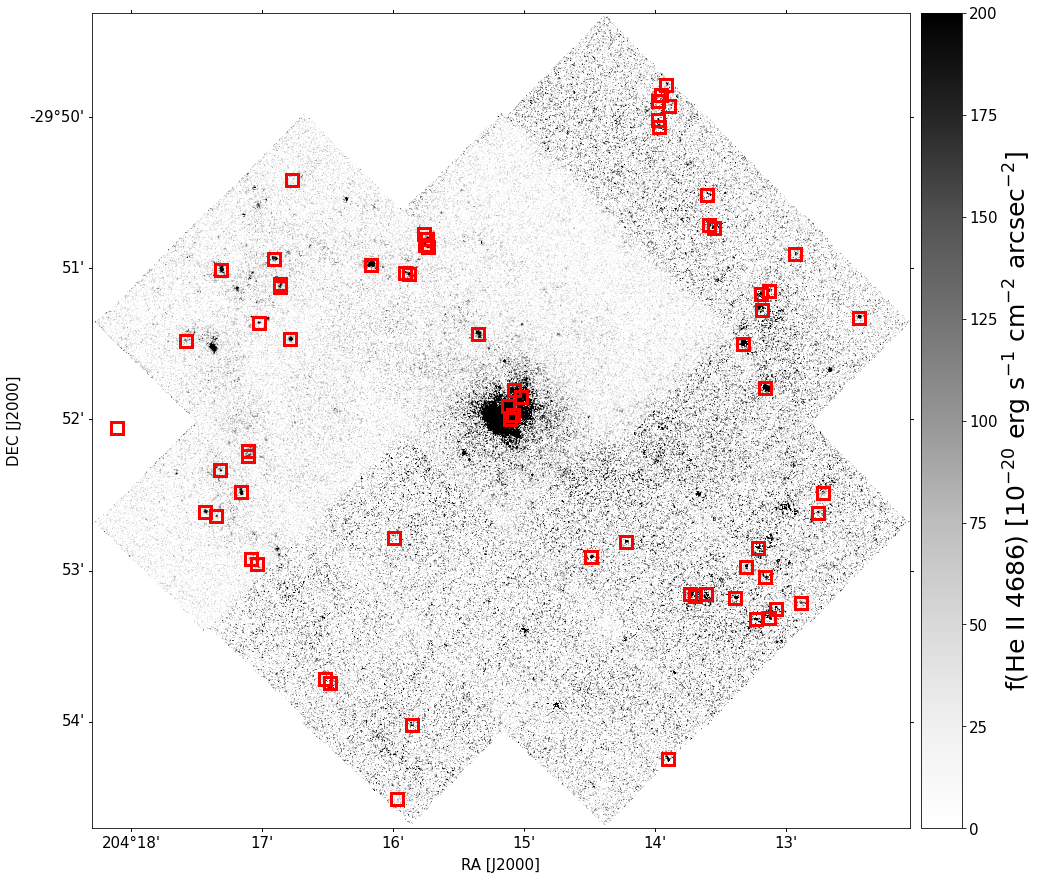}
    \caption{Location of the identified WR on a map of \ion{He}{ii}~$\lambda$4686. The source that appears out of the MUSE FoV was observed in the non-extended wavelength mode. It was extracted as part of the \citet{hadfield05} catalogue and, despite the lack of \ion{He}{ii}~$\lambda$4686 emission, featured emission lines characteristic of a WR.}
    \label{fig:WR_cc}
\end{figure*}

\clearpage
\onecolumn
\topcaption{Coordinates of the confirmed WR candidates and cross-match with the catalogue of \citet{hadfield05}.}
\tablefirsthead{ \toprule \toprule
Obj. ID & Ra (J2000) & Dec (J2000) & Hadfied ID & Spectral classification\\ \midrule }
\tablehead{
\toprule \toprule
Obj. ID & Ra (J2000) & Dec (J2000) & Hadfied ID & Spectral classification\\ \midrule }
\begin{supertabular}{lllll}
WR1 & 13:37:03.4358 & -29:54:01.463 &  & WCL\\
WR2 & 13:36:52.9334 & -29:53:19.649 & & WCL\\
WR3 & 13:36:52.5156 & -29:53:18.951 & & WCL\\
WR4 & 13:36:52.3236 & -29:53:15.497 & 38 & WCL\\
WR5 & 13:36:51.5489 & -29:53:13.031 &  35, 36 & WCL \\
WR6 & 13:36:53.5601 & -29:53:11.078 &  & WCL\\
WR7 & 13:36:54.9309 & -29:53:09.485 &  & WCL\\
WR8 & 13:36:54.4484 & -29:53:09.685 &  & WC \\
WR9 & 13:36:52.6505 & -29:53:02.695 & 40 & WCE \\
WR10 & 13:36:53.2367 & -29:52:58.668  & & WCL\\
WR11 & 13:37:08.1649 & -29:52:57.743  & & WNL \\
WR12 & 13:37:08.3589 & -29:52:55.516  & 102 & WCE\\
WR13 & 13:36:57.9527 & -29:52:54.980  & 61 & WCL\\
WR14 & 13:36:52.8585 & -29:52:51.193  & 42 & WC\\
WR15 & 13:36:56.8925 & -29:52:48.897  & 59 & WNL \\
WR16 & 13:37:09.4049 & -29:52:38.767  & & WCE\\
WR17 & 13:36:51.0396 & -29:52:37.178  & 33 & WNL \\
WR18 & 13:37:09.7381 & -29:52:36.913  & 108 & WNL\\
WR19 & 13:37:08.6512 & -29:52:29.098  & 105 & WCL\\
WR20 & 13:37:09.2929 & -29:52:20.418 &  & WCE\\
WR21 & 13:37:08.4468 & -29:52:14.703 &  & WCE\\
WR22 & 13:37:08.4489 & -29:52:12.666  & 103 & WCE\\
WR23 & 13:36:52.6444 & -29:51:47.860  & 41 & WCL\\
WR24 & 13:36:53.3294 & -29:51:30.171  & 44 & WCL + WNL\\
WR25 & 13:37:10.3164 & -29:51:28.992  & 109 & WCL\\
WR26 & 13:37:07.1428 & -29:51:28.489 & & WNL + WCE\\
WR27 & 13:37:01.4173 & -29:51:26.391 & 74 & WCL\\
WR28 & 13:37:08.1053 & -29:51:22.042 &  & WNL \\
WR29 & 13:36:49.7945 & -29:51:19.780  & 25 & WNL \\
WR30 & 13:36:52.7543 & -29:51:16.855 &  & WNL\\
WR31 & 13:36:52.7781 & -29:51:10.482 &  & WCL\\
WR32 & 13:36:52.5184 & -29:51:09.191  & 39 & WCE\\
WR33 & 13:37:07.4636 & -29:51:07.716 &  & WCL\\
WR34 & 13:37:07.4568 & -29:51:06.347  & 97 & WCL\\
WR35 & 13:37:03.5195 & -29:51:02.594 &  & WCL\\
WR36 & 13:37:09.2508 & -29:51:00.797  & & WCL\\
WR37 & 13:37:03.6487 & -29:51:02.025 &  & WCL\\
WR38 & 13:37:04.6779 & -29:50:58.724 & 86 & WCL \\
WR39 & 13:37:07.6316 & -29:50:56.561 &  & WNE\\
WR40 & 13:36:51.7288 & -29:50:54.711 &  & WNL \\
WR41 & 13:37:02.9289 & -29:50:51.907 &  & WCL\\
WR42 & 13:37:03.0393 & -29:50:50.864  & 78 & WCL \\
WR43 & 13:37:02.9674 & -29:50:48.732 &  & WCE\\
WR44 & 13:36:54.1999 & -29:50:44.021 &  & WCL\\
WR45 & 13:36:54.3722 & -29:50:43.003 &  & WNL\\
WR46 & 13:36:54.4221 & -29:50:31.197 &  & WCE\\
WR47 & 13:36:55.8911 & -29:50:03.944 &  & WCE\\
WR48 & 13:36:55.9245 & -29:50:01.499 &  & WCE\\
WR49 & 13:36:55.5874 & -29:49:55.766 &  & WCE\\
WR50 & 13:36:55.9334 & -29:49:53.779  & 57 & WNL \\
WR51 & 13:36:55.8305 & -29:49:51.351 &  & WCE\\
WR52 & 13:36:55.6703 & -29:49:47.354 &  & WNL\\
WR53 & 13:37:03.9700 & -29:52:47.500 & 82 & WCL\\
WR54 & 13:37:06.1000 & -29:53:43.500  & 90 & WCL\\
WR55 & 13:37:07.1000 & -29:50:25.200  & 94 & WCL\\
WR56 & 13:37:00.3306 & -29:51:48.669 &  & WCL\\
WR57 & 13:37:00.1008 & -29:51:51.273 &  & WCL\\
WR58 & 13:37:00.4937 & -29:51:55.034 & & WCL\\
WR59 & 13:36:55.6282 & -29:54:15.149 &  & WCE\\
WR60 & 13:37:05.9271 & -29:53:44.863 &  & WCL\\
WR61 & 13:36:54.7982 & -29:53:10.488 & & WCL\\
WR62 & 13:36:50.8675 & -29:52:29.262 &  & WNE\\
WR63 & 13:37:03.0770 & -29:50:46.687 & & WCL \\
WR64 & 13:36:54.1999 & -29:50:44.021 &  & WCL\\
WR65 & 13:37:03.8800 & -29:54:31.100 & 81 & WCL\\
WR66 & 13:37:12.4500 & -29:52:03.700  & 110 & WCL\\
WR67 & 13:37:00.4270 & -29:51:59.953 &  & WCE\\
WR68 & 13:37:00.3380 & -29:51:58.314 & & WCL\\
\end{supertabular}
\label{table:WR_cc}
\twocolumn
\clearpage

\end{appendix}

\end{document}